%% file: main.tex
\documentclass[acmsmall]{acmart}

\AtBeginDocument{%
  }

\usepackage{listings}

\begin{document}
\setcopyright{cc}
\setcctype{by}
\acmJournal{PACMHCI}
\acmYear{2026} \acmVolume{10} \acmNumber{6} \acmArticle{CSCW059}
\acmMonth{10} \acmDOI{10.1145/3816907}
\received{May 13, 2025}
\received[revised]{January 13, 2026}
\received[accepted]{April 9, 2026}

\title{Bringing Everyone to the Table: An Experimental Study of LLM-Facilitated Group Decision Making
}

\author{Mohammed Alsobay}
\email{malsobay@microsoft.com}
\affiliation{%
  \institution{Microsoft Research}
  \city{New York}
  \state{NY}
  \country{USA}
}

\author{David M. Rothschild}
\affiliation{%
    \institution{Microsoft Research} 
    \city{New York} 
    \state{NY}
    \country{USA}}
\email{davidmr@microsoft.com}

\author{Jake M. Hofman}
\affiliation{%
  \institution{Microsoft Research}
  \city{New York}
  \state{NY}
  \country{USA}}
\email{jmh@microsoft.com}

\author{Daniel G. Goldstein}
\affiliation{%
  \institution{Microsoft Research}
  \city{New York}
  \state{NY}
  \country{USA}}
\email{dgg@microsoft.com}

\renewcommand{\shortauthors}{Alsobay et al.}

\begin{abstract}
  Group decision-making often suffers from uneven information sharing, hindering decision quality. While large language models (LLMs) have been widely studied as aids for individuals, their potential to support groups of users, potentially as facilitators, is relatively underexplored. We present a pre-registered randomized experiment with 1,475 participants assigned to 281 live groups completing a hidden profile task—selecting an optimal city for a hypothetical sporting event—under one of four facilitation conditions: no facilitation, a one-time message prompting information sharing, a human facilitator, or an LLM (GPT-4o) facilitator. We find that LLM facilitation increased information shared within a discussion by raising the minimum level of engagement with the task among group members, and that these gains came at limited cost in terms of participants' attitudes towards the task, their group, or their facilitator. Whether by human or AI, there was no significant effect of facilitation on the final decision outcome, suggesting that even substantial but partial increases in information sharing were insufficient to overcome the hidden profile effect studied. To support the design and evaluation of LLM-mediated group decision-making systems, we release our data and our experimental platform, the Group-AI Interaction Laboratory (GRAIL), as an open-source tool.

\end{abstract}

\ccsdesc[500]{Human-centered computing~Empirical studies in collaborative and social computing}

\keywords{decision making, human-AI interaction, collective intelligence, experiments}

\maketitle

\input{contents/intro}
\input{contents/related_work}
\input{contents/methods}
\input{contents/results}
\input{contents/discussion}
\input{contents/conclusion}

\input{contents/acknowledgments}

\bibliographystyle{ACM-Reference-Format}
\bibliography{main}

\input{contents/appendix}

\end{document}

%% file: contents/intro.tex
\section{Introduction}
From tribal councils to committee meetings, group decision-making is a cornerstone of human societal behavior across eras and cultures. Groups are assembled to make decisions in critical functions such as hiring \cite{Tavana1993-pq}, patient care \cite{DiPierro2022-lw}, and crisis response \cite{Li2022-gb}, and factors moderating the performance of groups have been studied extensively \cite{Larson1993-ul, Larson2010-mo, Kozlowski2006-pb, Almaatouq2021-nf, McGrath1984-gp, Hackman1975-yi}. Although group-based tasks are ubiquitous in work environments, groups face challenges such as social loafing \cite{Latane1979-kb}, groupthink \cite{Baron2005-ff}, and polarization \cite{Isenberg1986-tk}, as well as various biases in information sharing, discussion, and evaluation \cite{Stasser1985-em, Mojzisch2010-ol,Schulz-Hardt2016-og}. 

Many efforts have been made to measure group aptitude and to design interventions that improve group performance. One such class of interventions involves the use of algorithms to influence collective behavior, ranging in sophistication from simple, noisy bots that improve coordination in networked groups \cite{Christakis2008-xv}, to graph neural networks that rewire human networks to promote cooperation \cite{McKee2023-hn}. With the launch of ChatGPT in November 2022, large language models (LLMs) have entered the public lexicon and become the focus of research exploring how these models can augment human abilities. While a broad body of literature has explored how LLMs can assist \textit{individual} users in a wide range of domains, including writing, programming, art, and education \cite{Noy2023-pq, Mozannar2024-qh, Wu2021-eb, Kumar2023-uc}, our current understanding of how these models can assist \textit{groups} of people using them together is relatively nascent \cite{Chiang2024-cz,Lee2024-jy,Mao2024-tb, Vanukuru2025-od}. 

Recent work has demonstrated the ability of LLMs to help groups achieve consensus \cite{Tessler2024-hq, Small2023-qg} and discuss divisive topics \cite{Argyle2023-st} through static interventions (i.e., summaries or suggestions delivered to group members). In contrast to post hoc tools such as meeting summaries, integrating LLMs as active participants in discussions offers the opportunity to support groups and increase discussion quality in real-time, by increasing desirable outcomes such as intentionality \cite{Chen2025-je} and inclusion \cite{Houtti2025-eh}. More specifically, as meeting facilitators, LLMs may excel in aspects of facilitation that would be taxing for their human counterparts, particularly in managing discussions that proceed at a high volume and rapid pace and keeping track of who has and has not participated. In this direction, recent work has addressed the design principles underlying such active LLM-based interventions \cite{Mao2024-tb} and empirically studied the effect of LLM-based interventions such as dissent on group decision-making and design processes \cite{Chiang2024-cz, Lee2024-jy}. Together, these studies point to a growing design space of LLM facilitation strategies that vary in their timing, target, and interaction style, and in doing so, shape how groups coordinate and deliberate.

In this work, we study the role of LLM-based facilitators in decision-making discussions in which critical information is asymmetrically distributed among group members, requiring effective information sharing to identify the optimal solution. In such settings, groups often fail to surface and effectively integrate uniquely held information, due to a combination of factors including uneven participation, conversational dynamics, and biases in how information is attended to and interpreted. These dynamics can lead to suboptimal decisions, particularly when key information is known to only a subset of participants, or even a single individual. 

We introduce LLMs as facilitators to aid groups completing ``hidden profile'' tasks (HPTs) \cite{Stasser1985-em}, a well-studied experimental paradigm of decision-making under asymmetric information. By enacting structural interventions that support participation and information synthesis, LLMs may be effective aids to groups facing such decisions. To further our understanding of LLM facilitation in this setting, we ask the following research questions (RQs):

\begin{itemize}
    \item \textbf{RQ1:} Relative to no facilitation, written guidance, and human facilitation, does LLM facilitation increase information sharing in group decision-making under asymmetric information? 
    \item \textbf{RQ2:} Beyond the volume of information shared, how does LLM facilitation affect the process of information aggregation, and the quality of the decision made? 
    \item \textbf{RQ3:} How do participants subjectively evaluate LLM-facilitated experiences of group decision-making? 
\end{itemize}

To investigate these questions, we conducted a pre-registered randomized experiment (\href{https://aspredicted.org/23df-gtys.pdf}{AsPredicted \#192061}) in which 281 groups of five decision makers completed a hidden profile task and were randomly assigned to complete the task either unassisted or with the help of a human facilitator or an LLM facilitator. We find that LLM facilitation increased the volume and breadth of information shared, without negatively impacting self-reported attitudes of group cohesion and productivity. However, despite the increased information sharing, groups still made suboptimal decisions, emphasizing the strength of the biases at play and motivating research into algorithmic facilitation tools that are targeted at overcoming well-documented biases in group discussion and decision making. In support of future research, we also release the full experimental dataset and the Group-AI Interaction Laboratory (GRAIL), an open-source interface for studying group-AI interaction in conversational tasks.

%% file: contents/related_work.tex
\section{Related Work}
In this work, we study the impact of LLM facilitation on group decision making under asymmetric information. Our research is motivated by the well-documented biases and communication failures exhibited by groups in settings of asymmetric information, and builds upon previous work in the design of group-AI interaction and group decision support technologies. We contribute to these lines of research by complementing existing studies with large-scale, pre-registered experimental evidence describing the impact of LLM facilitation on group behavior while completing a challenging hidden profile task. 

\subsection{The Hidden Profile Task Paradigm of Group Decision-making Research}
Since the seminal work of Stasser and Titus \cite{Stasser1985-em}, hidden profile tasks (HPTs) have been used extensively to study the dynamics of decision making in groups whose members have asymmetric information  \cite{Stasser2003-kw, Lu2012-hh, Sohrab2015-bs, Reimer2010-no, Brodbeck2007-ae}. In an HPT, a group of decision makers is tasked with selecting one of several predefined options, with information about the options  distributed among different group members. Each group member receives a distinct information set (or ``profile'') about the various options, typically delivered as a list of facts about each option. The visibility (public/shared/private) and polarity (positive/negative/neutral) of pieces of information and their distribution among group members is deliberately structured such that individually accessible information suggests one alternative as optimal, while the full information set would identify a different option as superior. This information asymmetry creates a decision scenario where optimal group performance requires effective information exchange and integration rather than simple majority voting or averaging of individual preferences. 

One familiar with the literature on the wisdom of crowds \cite{Surowiecki2005-qv} and collective intelligence \cite{Woolley_undated-rg} may intuitively expect groups to outperform individuals in decision making tasks, including HPTs. However, prior research finds that groups typically do not exhaustively share information due to effort and time constraints \cite{Stasser1985-em}. The distribution of information under the HPT induces two key information sharing biases that hinder group progress. In groups completing an HPT, shared information tends to be surfaced more frequently than private information (shared information bias \cite{Nicholson2021-gk, Larson1996-zt}), and participants are more likely to share information that supports the option that is appealing under their personal information set (individual preference bias \cite{Kelly1999-wk, Schulz-Hardt2016-og}). 

More broadly, these dynamics are amplified by well-documented group processes. For example, social loafing \cite{Latane1979-kb} can reduce participation, increasing the chance that privately held information is never shared; communication apprehension \cite{McCROSKEY1977-sm, Nicholson2021-gk} can discourage individuals from contributing unique or dissenting information; and groupthink \cite{Baron2005-ff} can lead groups to converge prematurely on an initially favored option without fully considering alternatives. Collectively, these factors contribute to a central challenge of the HPT paradigm: groups often fail not because the relevant information is unavailable, but because it is not effectively surfaced and integrated during discussion.

\subsection{Computational Approaches to Group Facilitation}
Broadly put, the role of a facilitator is to enact process and content interventions that help a group achieve its objectives \cite{Viller1991-uz, Miranda1999-bd}. \textit{Process interventions} introduce and reinforce structure in a group's discussion, such as imposing an agenda or by managing specific discussion formats (e.g., the Delphi method and the nominal group technique \cite{Fink1984-bx}). In conversation, facilitators enacting process interventions typically perform actions such as timekeeping, managing participation, and supporting synthesis, without introducing their own substantive judgments or domain expertise. In contrast, \textit{content interventions} involve the facilitator contributing their own judgment, interpretations, or recommendations to the discussion at hand, such as advocating for a particular option or introducing new information and perspectives.  

Computational approaches to facilitation build on a rich body of work in the CSCW and broader HCI communities on group decision support systems (GDSSs) \cite{DeSanctis1987-oh} and conversational agents (CAs) \cite{Diederich2022-yj}. In 1993, Phillips \& Phillips argued that, while computers could support facilitation as a medium for interaction, ``the key functions of the facilitator, observing, attending, maintaining awareness of feelings, and intervening, cannot, of course, be carried out by computers...'' \cite{Phillips1993-fb}. Since then, computational approaches to these functions have arguably progressed to the level of practical utility, and subsequent work has explored technological solutions to enact the structural and conversational roles of the facilitator in group settings. For example, Kim et al.\ \cite{Kim2021-qw} demonstrate the utility of CAs that interactively add structure (e.g., turn-taking) to group debates, while Do et al.\ \cite{Do2022-am} show how agents can sensitively provide feedback to underperforming group members, and Shamekhi et al.\ \cite{Shamekhi2018-gs, Shamekhi2019-ht} examine how embodiment influences perceptions of rapport, trust, intelligence, power, and efficacy of facilitative agents in decision-making groups.

In the context of decision making under asymmetric information, process interventions may, in theory, help groups avoid suboptimal decisions. In hidden profile tasks, group members collectively possess the information required to identify the optimal choice, yet this information is often unevenly shared and insufficiently integrated into the final decision. By eliciting contributions from all members and structuring how information is shared and synthesized, process interventions are well positioned to mitigate these failures of information exchange. This makes hidden profile tasks a natural setting in which to study the effects of process-oriented facilitation. At the same time, while such interventions may be beneficial in theory, it remains an empirical question whether LLM-based systems can implement such facilitation effectively in practice.

Effective LLM facilitators may allow us to easily conduct these facilitatory functions with greater fidelity and scale, and across a wide range of domains, potentially broadening access to facilitative support. Professional facilitators, such as those found on \href{https://findafacilitator.com/}{Find a Facilitator}, command fees in the range of \$1,500-\$3,000/day\footnote{Interquartile range of 316 facilitators in the United States found by searching for the ``meeting facilitation'' service on Find a Facilitator (https://findafacilitator.com/), as of February 27th, 2025.}, which may be out of reach for groups in a wide range of decision making settings. The comparatively low (and decreasing) marginal cost of LLM-based systems may offer a more scalable and cost-effective alternative.  

\subsection{Group-AI Interaction}
A broad range of algorithms have been used to influence group behavior with the goal of overcoming detrimental group processes, ranging in sophistication from simple, noisy bots that improve coordination in networked groups \cite{Christakis2008-xv}, to graph neural networks that rewire human networks to promote cooperation \cite{McKee2023-hn} and variational autoencoders used to influence group dynamics in a hidden profile task \cite{Pescetelli2021-nk}. 

More recently, large language models (LLMs) have been explored as a tool to influence group dynamics and outcomes in a range of settings. One form of such interventions is characterized by static or post-hoc LLM output provided to groups; in these applications, the LLM serves as a tool that provides input to the group in a highly structured format, rather than as an active participant in a group's interactions. Examples of such interventions include Tessler et al.'s \cite{Tessler2024-hq} LLM-based ``Habermas Machine'' which iteratively generates consensus statements for social and political issues, as well as approaches that use LLMs to help users phrase their messages to increase the sense of feeling understood in divisive conversations \cite{Argyle2023-st} and improve out-group cooperation \cite{Claggett2025-ov}. 

As Mao et al.\ \cite{Mao2024-tb} and Do et al.\ \cite{Do2022-am} emphasize, transitioning from structured forms of support to an interactive, multi-user chat setting expands the design space from \textit{what} to say, to include the \textit{when} and \textit{who}. To address the added complexity of multi-party interactivity in chat-based settings, frameworks such as those proposed by Mao et al.\ \cite{Mao2024-tb} and Papachristou et al.\ \cite{Papachristou2025-ex} introduce multi-component LLM systems in which each component tackles a single facet of interaction (e.g. timing, content, recipient selection, intent extraction, and coordination), while the ``Observe, Ask, Intervene'' framework proposed by Houtti et al.\ \cite{Houtti2025-eh} focuses on the process by which such interactive interventions should be delivered. 

Within the context of meetings and group discussions, LLMs have been explored as ``devil's advocates'' that introduce disconfirmatory opinions \cite{Chiang2024-cz}, as coordinators of meeting scheduling processes \cite{Papachristou2025-ex}, and as tools that support temporal work across recurrent meetings \cite{Vanukuru2025-od}. We build upon and extend this literature by conducting a pre-registered, large-scale (281 groups), highly interactive (5-6 members per group) experiment studying the role of single LLMs as decision-making facilitators in a setting that is particularly challenging due to its asymmetric and adversarial information landscape.

%% file: contents/methods.tex
\section{Methods}
\subsection{Participants}
In this pre-registered experiment (\href{https://aspredicted.org/23df-gtys.pdf}{AsPredicted \#192061}), 1,475 participants were recruited from Prolific and randomly assigned into groups of five decision makers (with a sixth participant added as facilitator in one treatment condition). All participants resided in the United States and the mean age was 39 years (SD = 12.5), with 54\% identifying as male and 45\% as female (the remainder declined to answer).

Participants were compensated a flat amount of \$3.75 for full completion of the study, and the median hourly compensation rate was \$14.40/hr. The study was approved by the Microsoft Research Institutional Review Board (MSR IRB; Record \#10883), and all participants provided explicit consent to participate in this study. For recruitment materials, please see Appendix \ref{app:recruitment}.  

\subsection{Experiment Design \& Implementation}
\subsubsection{Facilitation}
The focal manipulation of this study was the availability of facilitative support to assist groups during deliberation. Each group was randomly assigned to receive one of four facilitation treatments: 

\begin{enumerate}
    \item \textbf{No facilitation} (``None'', N=70 groups), in which a group completed the task without any facilitative support.
    \item \textbf{A one-time message} (``Message'', N=71 groups), in which the group was shown a message saying ``People may have different information about what is being discussed in this meeting, so encourage everyone to share all of the relevant information they have.'' This notice was shown at the beginning of the group's deliberation period, and gradually disappeared once the discussion began. As a minimal pre-deliberation prompt, this intervention mirrors active control conditions previously studied by Stasser et al. \cite{Stasser1995-gs, Stasser2000-yo}.
    \item \textbf{Human facilitation} (``Human'', N=70 groups), in which a sixth human participant was added to the group to act as a facilitator. This participant did not receive any private task information, and therefore operated as a neutral, external lay facilitator focused on structuring the group's discussion rather than contributing substantively to the decision.    
    \item \textbf{LLM facilitation} (``LLM'', N=70 groups), in which an LLM (GPT-4o) acted as a facilitator.
    
\end{enumerate}

Neither the human nor LLM facilitator was provided with information relevant to any of the options in the hidden profile task, and therefore both functioned as external participants whose role was limited to structuring the group's deliberation rather than contributing task-relevant content. Both facilitators were presented with the same core prompt to ``first, make sure that everyone is heard from and shares what they know and, second, act as a scoreboard and keep track of pros and cons'', and were also given the reminder presented in the ``Message'' facilitation condition (in the ``Human'' and ``LLM'' conditions, the message was only provided to the facilitator, not the participants). If the group was assigned to have an LLM facilitator, the LLM sent a message to the group every 90 seconds, in addition to responding every time it was directly addressed by the group; each request sent to the LLM includes the full timestamped transcript preceding the request, as well as general information about meeting objectives, attendees, and time elapsed/remaining. For implementation details of each facilitative condition, including screenshots, please see Appendix \ref{app:interface-screenshots}. The implementation of the LLM facilitator and its prompt structure are further detailed in Appendix \ref{app:facilitator-implementation}.\looseness=-1

\subsubsection{The Hidden Profile Task}
In this experiment's hidden profile task, groups were tasked with selecting one of three fictional cities---Eldoron, Myloria, and Cragnio---to host an international sporting event akin to the Olympics. Each city was described by 10 different facts, and facts had five categories: weather, tourism, transportation, community, and a miscellaneous category for neutral information that did not plausibly influence the decision to host a sporting event. Each fact about a city had a positive (e.g., ``friendly and welcoming residents''), negative (e.g., ``higher than average humidity''), or neutral valence (e.g., ``considered a hub for biotech in the region''), validated through a pre-experiment survey. The structure of the task, including group size, number of options, and information architecture, was modeled after the ``Grogan Air'' hidden profile task \cite{AmesUnknown-vy}, with the content modified for the international sporting event context. For details of the hidden profile task design and valence measurement survey, please see Appendix \ref{app:hpt-design}. 

Under full information and a unit-weight linear model (UWLM) of utility \cite{Gigerenzer1996-tu} which assigns a score of 1 to positive facts, 0 to neutral facts, and -1 to negative facts, Eldoron is the most favorable option with a total score of 6, followed by Myloria with a score of 2, and Cragnio with a score of 0. However, each group member only had access to a subset of facts; facts could either be public (available to all group members), shared (available to multiple members), or private (available to only one member). To obscure the utility of Eldoron and create the ``hidden'' profile, the polarity (positive/negative/neutral) and visibility (public/shared/private) of the facts were manipulated such that Myloria was the most appealing option in each individual information set, as detailed in Table \ref{tab:info_sets}. For the full list of facts and the distribution of information among the participants, please see Figure \ref{fig:hpt-info} in the Appendix.

\begin{table}
    \centering
    \begin{tabular}{cccc}
        \textbf{\underline{Information Set}} & \textbf{\underline{Eldoron}} & \textbf{\underline{Myloria}} & \textbf{\underline{Cragnio}} \\
        
        Full & \textbf{6} & 2 & 0\\
        Blue & 2 & \textbf{3} & 1\\
        Green & 1 & \textbf{3} & 2\\
         Orange & 2 & \textbf{3} & 2\\
         Pink & 1 & \textbf{3} & 2\\
         Red & 1 & \textbf{3} & 2
    \end{tabular}
    \caption[Utility scores for each city (Eldoron, Myloria, Cragnio) by information set]{\textbf{Utility scores for each city (Eldoron, Myloria, Cragnio) by information set.} Under full information, Eldoron has the highest utility, whereas individual participant information sets (Blue, Green, Orange, Pink, Red) were manipulated to create a hidden profile favoring Myloria as the most attractive city. Utility scores reflect the sum of valence values assigned to positive (+1), neutral (0), and negative (-1) facts based on a unit-weight linear model (UWLM).}
    \label{tab:info_sets}
\end{table}

\subsubsection{Experimental flow}
\begin{figure}[h!]
    \centering
\includegraphics[width=0.85\linewidth]{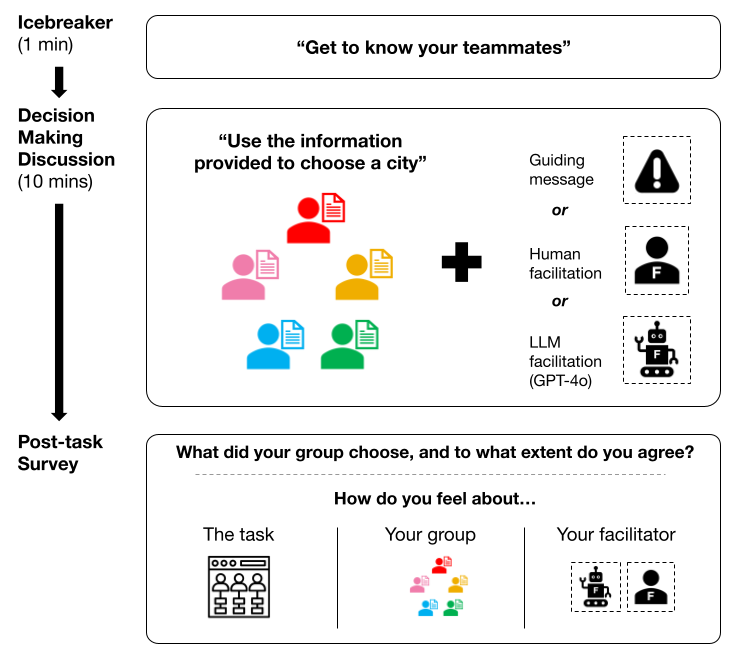}
    \caption[Experimental design]{\textbf{Experimental design.} After reading an overview of the task and completing a brief comprehension check, participants were divided into groups of 5 decision makers (and a 6th participant member acting as facilitator, if assigned one) to proceed through the 3 stages of the experiment. (A) First, participants completed a 1-minute icebreaker exercise in which they met their group using the same discussion interface they would use for the main task. (B) After the icebreaker, groups had 10 minutes to complete the hidden profile task, and were randomized to either a control condition with no facilitative support, or one of the three facilitative treatments: ``Message'', ``Human'', or ``LLM''. (C) After the main discussion, each participant completed an exit survey individually, in which they reported the group's decision and their subjective attitudes towards the task, their group, and the facilitator, if they were assigned one.}
    \label{fig:exp-design}
    \Description{Experimental flow diagram. Participants first completed a 1-minute icebreaker, then engaged in a 10-minute hidden-profile decision-making discussion in groups of five. During decision making, groups were randomly assigned to one of four facilitation conditions: no facilitation, a one-time message encouraging information sharing, a human facilitator, or an LLM facilitator. After the main discussion, participants completed an individual post-task survey reporting their group's decision and their subjective evaluation of the experience.}
\end{figure}

\begin{figure}[h!]
    \centering
\includegraphics[width=1\linewidth]{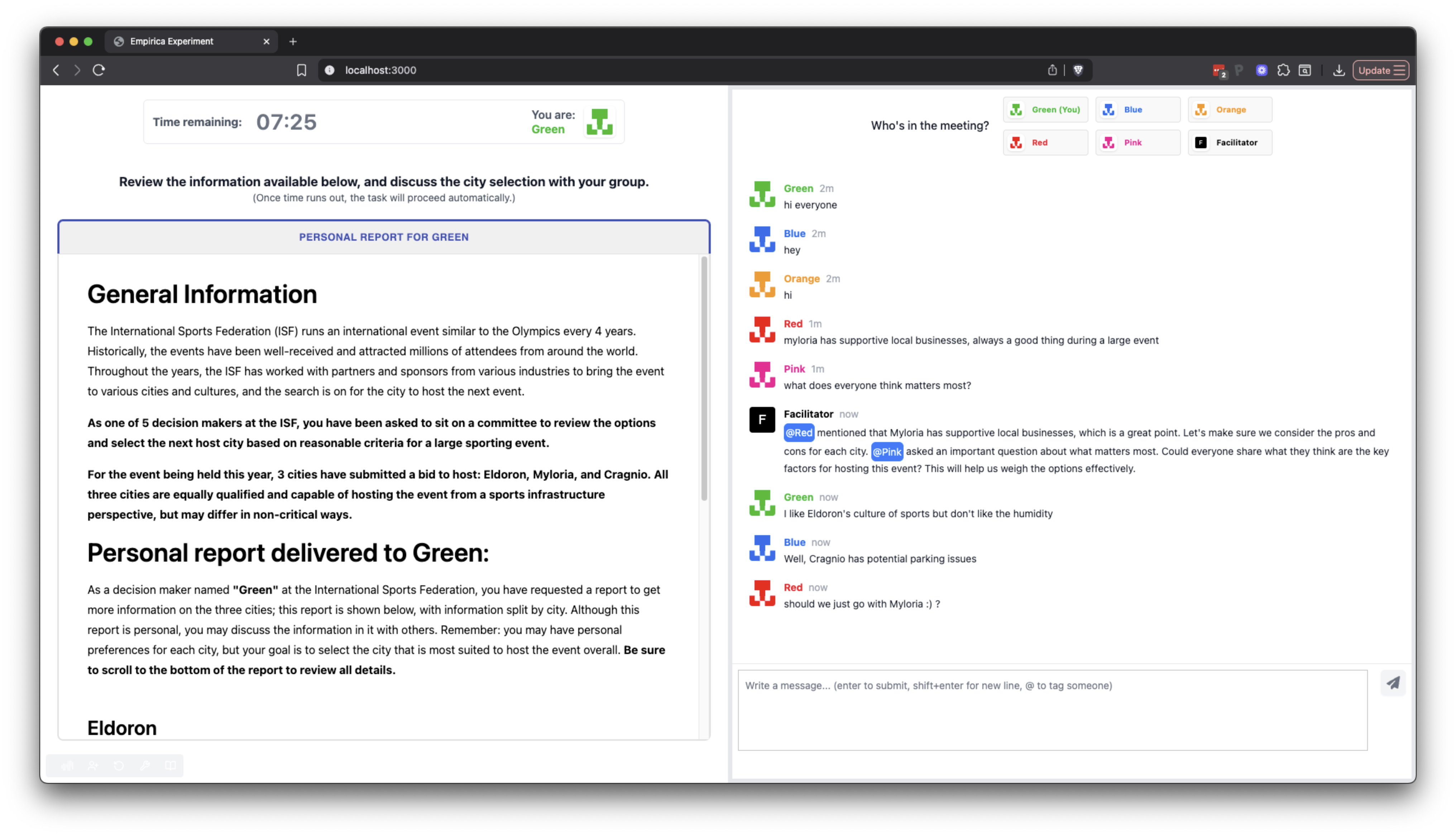}
    \caption[Experiment interface]{\textbf{Experiment interface.} On the left half of the screen, participants could review the general instructions and the unique set of information available to them about each of the 3 cities. On the right half, the group communicated through a shared chat window; participants could use ``@'' tags to directly address others, including the facilitator.}
    \label{fig:exp-ui}
    \Description{Screenshot of the experimental interface. On the left half of the screen, participants can review the general instructions and the unique set of information available to them about each of the 3 cities. On the right half, the group communicates through a shared chat window; participants can use ``@'' tags to directly address others, including the facilitator.}
\end{figure}

During onboarding, participants are informed of the purpose and duration of the meeting, the size of their group, and the availability and type of facilitator supporting their group, if they have been assigned one. After reviewing the instructions and completing a brief comprehension check, each group proceeds through three stages of the experiment, illustrated in Figure \ref{fig:exp-design}, using the chat interface illustrated in Figure \ref{fig:exp-ui}: 
\begin{enumerate}
    \item \textbf{Icebreaker:} Groups begin by spending 1 minute introducing themselves to each other, using the same chat interface they will use for the main decision making task. During this stage, participants gain familiarity with the interface, allowing them to immediately begin the decision-making discussion in the next stage. Participants are not shown any task-relevant information during this stage. While human facilitators may participate in the icebreaker stage, the LLM facilitator does not. 

    \item \textbf{Decision Making Discussion:} After the icebreaker stage, each participant is shown their personal information set, and the group is given 10 minutes to discuss the task. Groups must stay for the entirety of the 10 minutes. During this stage, facilitators in the ``Human'' and ``LLM'' conditions communicate with their group through the same chat window the group is using for their discussion. Instead of receiving task-specific information, as is shown to the five ``decision making'' participants, the human facilitator's interface shows the facilitator's instructions. 

    \item \textbf{Post-task Survey:} After the decision making discussion, all participants (including human facilitators) complete an exit survey individually. They are asked to report the decision made by the group, as well as their subjective evaluations of the task, their group, and their facilitator (if they were assigned one). For screenshots of the introduction and exit survey, please refer to Appendix \ref{app:intro-exit}.
\end{enumerate}

Within each group, participants are randomly assigned colors (``Blue'', ``Pink'', ``Green'', ``Orange'', ``Red'') as nicknames\footnote{We selected colors as nicknames for ease of use by participants and to avoid stronger implicit associations (e.g., by roles or human names), while acknowledging the possibility that color labels themselves could still influence the behavior of both participants and the LLM.}; across all groups, each color is associated with an information set (which includes all public information, and any shared or private information assigned to them), presented in ``personal briefs''. Through the interface shown in Figure \ref{fig:exp-ui}, participants can view their personal brief on the left side of the screen, and communicate through a chat window on the right side of the screen. The interface is configured to disable direct copying and pasting of the information presented in the personal brief, preserving the limitations on information sharing that make it necessary to choose which information to share, and preventing participants from easily sharing all available facts or copy-pasting the content into an external LLM to be summarized. 

\subsection{Outcome and Process Measures}
To answer RQ1, our pre-registered primary outcome is the volume of information covered within each group, operationalized as the total number of distinct facts mentioned during the group's discussion, ranging from 0 to a maximum of 30 total facts. To quantify this outcome, each message sent by participants is annotated by GPT-4o to identify all directly or indirectly mentioned facts (for further details, please refer to Appendix \ref{app:fact-detection}). Our focus on information coverage is motivated by the fact that information aggregation is a necessary (albeit insufficient, as we discuss in Section \ref{ms:final-decisions}) condition for effective decision-making not only in our instantiation of the HPT, but across a wide range of HPT designs and group performance evaluation schemes \cite{Mesmer-Magnus2009-fa}.

To investigate RQ2, which studies the impact of our facilitative treatments on the discussion dynamics, we measure several process outcomes. Discussion density, defined as the ratio of facts surfaced to the number of messages sent in the discussion (excluding facilitator messages), allows us to differentiate between an increase in fact sharing driven by longer conversations, as opposed to an increase driven by a different discussion focus. To understand the impact of facilitation on option exploration, we measure the Gini coefficient of the facts surfaced for each of the three options within a group, where a lower Gini coefficient implies more even fact coverage of the options. 

In settings of asymmetric information, such as the hidden profile task, individual engagement is critical; if a group member does not contribute to the discussion, key information may be locked out of the decision making process. Within each group, we measure the upper and lower bounds of individual engagement by measuring the minimum and maximum number of new facts added to the discussion by an individual group member. A distinct but related measure is whether a group achieves full participation, which we define as whether every member has surfaced at least one new fact not previously mentioned in the discussion.

Finally, we define measures describing the information surfaced, and the decision made by the group. Within each group, we define an option's ``normative score'' as the score that would be given to an option on the basis of the facts surfaced by the group, assuming fact polarity were perfectly perceived by the groups as intended, and use these scores to find the decision the group would make under a unit-weight linear model of decision making. We contrast this normative decision to the actual decision made by the group, defined as the option reported to be the group's decision by a majority of its members, excluding human facilitators. All 281 groups yielded a clear majority, with 19 groups having a 4-1 split of votes, 8 groups having a 3-2 split, and the remaining groups making unanimous decisions. 

RQ3 is motivated by the fact that, while a given treatment may lead to outcome or process improvements, such improvements could possibly come at a cost to the subjective experience of decision makers. To study this, we measure individual participants' subjective evaluation of three dimensions--the task, their group, and their facilitator, if they had one--in a post-task survey. Subjective task experience is measured through the NASA Task Load Index \cite{Hart1988-rg}, a 6-item instrument capturing perceived workload across mental, physical, and temporal demand, as well as performance, effort, and frustration.  

Following the TLX, the subjective evaluation of the group begins with a general, free-text question (``How did you feel about your group in the task?''), followed by four 5-point Likert scale questions intended to capture commonly-studied facets of group decision-making and discussion quality, namely participation and inclusion (``I feel that I was able to contribute to the group discussion''), agency (``I feel that I was able to influence the group's decision''), productivity (``The group's discussion felt productive''), and structure (``The group had a structured way of collecting and summarizing information to reach a decision''). The final item of the group evaluation intended to measure group cohesion, by asking a binary question of whether the participant would hypothetically prefer to repeat the task with the same group or a new, random group. 

Lastly, following the structure of the group evaluation, the facilitator evaluation begins with a similarly phrased free-text question about the facilitator, followed by four 5-point Likert scale questions to evaluate the performance of the facilitator, along dimensions of information elicitation (``The facilitator encouraged the group to share information''), information synthesis (``The facilitator helped the group summarize the information shared to reach a decision''), task focus (``The facilitator was able to keep the group focused on the task''), and distraction and process interruption (``The facilitator was distracting''). The facilitator evaluation concludes by asking ``You completed the task with [no, a human, an AI] facilitator. If you were to repeat the task, which type of facilitation would you prefer?'', with the available options being ``None'', ``Human'', and ``LLM''. 

For analysis, TLX and Likert-based responses are aggregated to the group level by taking the mean of individual non-facilitator responses within each group, aligning the unit of analysis with that of treatment assignment. Subjective survey completion rates exceeded 98\% for all TLX items, 95\% for all group and facilitator Likert and multiple-choice items, and 68\% for optional free-text responses.

\subsection{Statistical Analyses and Data Processing}
As pre-registered (\href{https://aspredicted.org/23df-gtys.pdf}{AsPredicted \#192061}), we investigate RQ1 through confirmatory, pairwise, one-sided, independent-samples t-tests of the difference in mean fact coverage between experimental treatments, with sample size chosen based on pilot study data to achieve 70\% power in detecting a difference of 2 facts shared at a significance level of 5\%. In addition to the p-values for these pre-registered analyses, we also report adjusted p-values using the Benjamini-Hochberg procedure \cite{Benjamini1995-tq} as a conservative robustness check. All other analyses in this work are exploratory and reported with unadjusted p-values, and all subsequent t-tests are two-sided. In our exploratory analyses, we employ one-way analysis of variance (ANOVA) and Pearson's chi-squared test to test for differences between treatments in continuous and binary outcomes, respectively. 

To probe the relative weight of facts in a group's decision, we estimated a multinomial logit model of option selection as a function of normative UWLM scores. Let $Y_g$ denote the option selected by group $g$, and let $S_{g,E}$, $S_{g,M}$, and $S_{g,C}$ denote the surfaced normative scores for Eldoron, Myloria, and Cragnio, respectively. For each non-baseline option $j \in \{\text{Eldoron}, \text{Cragnio}\}$, we specify:
\[
\log \frac{\Pr(Y_g = j)}{\Pr(Y_g = \text{Myloria})}
=
\beta_{jE} S_{g,E}
+ \beta_{jM} S_{g,M}
+ \beta_{jC} S_{g,C}
\]
We set Myloria as the reference category, so each coefficient $\beta_{jk}$ captures the association between the surfaced normative score of option $k$ and the log-odds of choosing option $j$ (relative to Myloria), holding other scores constant. Importantly, because normative scores are themselves shaped by the experimental treatment, these estimates should be interpreted descriptively: they capture how surfaced information and decisions co-vary across groups, rather than isolating the causal effect of additional information holding the decision process fixed. As such, the coefficients reflect a combination of how groups weight information and how treatments jointly influence both information sharing and decision-making.

\subsubsection{Data processing}
Given the sensitivity of the decision-making task to private information that may be lost to a group if a participant leaves the study, our pre-registered inclusion criteria are that each participant sends at least one message, the group collectively sends at least 10 messages, and that all participants submit the task completion code (displayed at the end of the study) on Prolific. Groups were sampled from Prolific until the pre-registered count of 70 valid groups per experimental condition was met. During data collection, 7 groups in each of the ``None'', ``Message'', and ``LLM'' conditions and 15 groups in the ``Human'' condition failed to satisfy the inclusion criteria; one additional group was sampled in the Message condition as a consequence of ensuring equal-probability assignment across treatments. While the difference in exclusion rates is not statistically significant ($\chi^2(3,N=317)$ = 4.56, p = 0.21), the higher exclusion rate in the ``Human'' condition may reflect its larger group size (6 vs. 5), which mechanically increases the probability that at least one participant causes a group's exclusion. For more details on pre-registration, please refer to Appendix \ref{app:prereg}.

%% file: contents/results.tex
\section{Results}
\subsection{LLM facilitation increased information sharing in group discussions under asymmetric information}
As illustrated in Figure \ref{fig:info-sharing}A, we find that neither the one-time message treatment ($\Delta$ = 1.32, t(139) = 1.52, p = 0.066, Benjamini-Hochberg adjusted p = 0.082) nor human facilitation ($\Delta$ = 0.43, t(138) = 0.51, unadjusted and B-H adjusted p = 0.31) significantly increased the average number of facts shared relative to no facilitation. However, we find that LLM facilitation increased information sharing relative to all other conditions: $\Delta$ = 2.9, t(138) = 3.58, p $<$ 0.001 , B-H adjusted p = 0.001 for ``None''; $\Delta$ = 1.58, t(139) = 2.1, p = 0.021, B-H adjusted p = 0.035 for ``Message''; and $\Delta$ = 2.47, t(138) = 3.38, p $<$ 0.001 , B-H adjusted p = 0.001 for ``Human''. To contextualize the observed increase in information coverage with LLM facilitation relative to no facilitation (``LLM'' vs. ``None''), the effect size (Cohen's $d$) is 0.61, indicating a moderate-to-large practical effect of LLM facilitation on information coverage. 

\subsection{Process-level effects of LLM facilitation on group discussion and information sharing}
\begin{figure}[t!]
    \centering
\includegraphics[width=1\linewidth]{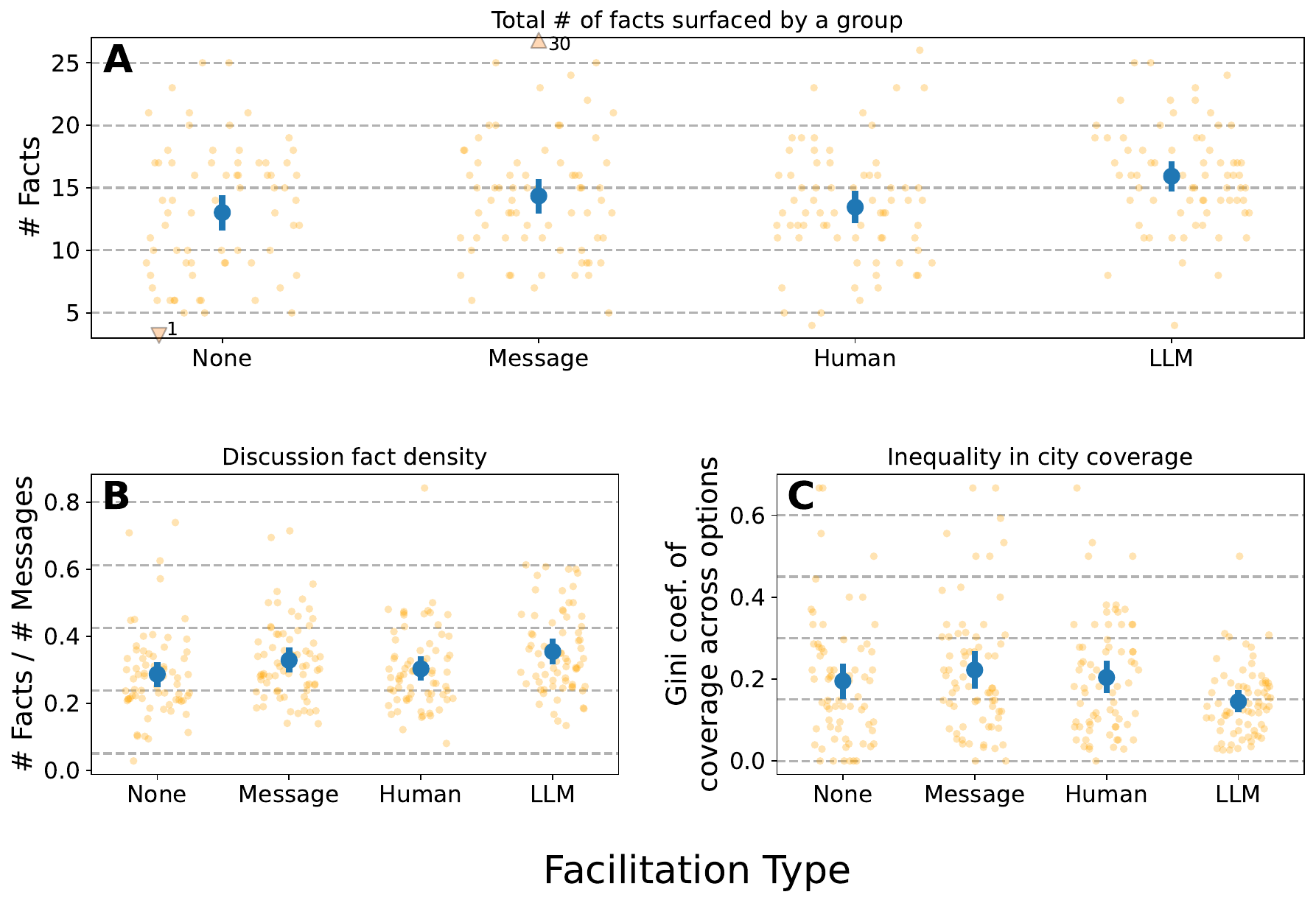}
    \caption[LLM facilitation increased fact sharing volume, density, and breadth during discussion]{\textbf{LLM facilitation increased fact sharing volume, density, and breadth during discussion.}\\(A) shows the total number of facts surfaced by a group. (B) shows the density of facts in a conversation (\# facts surfaced / \# messages, excluding facilitator messages). (C) shows the Gini coefficient of each group's covered information across each of the three options. Error bars indicate 95\% confidence intervals, and may be smaller than the marker. Scatter markers each represent a single group. For readability, border markers in panel (A) indicate points out of the y-axis bounds.}
    \label{fig:info-sharing}
    \Description{LLM facilitation increased the total amount of information shared, increased information density (facts per message), and led to more balanced discussion across the three candidate cities compared with the other facilitation conditions.}
\end{figure}

Despite LLM facilitators eliciting more facts from groups, we detect no significant difference in the average number of overall messages sent by the group members (excluding the facilitators) between treatments (F(3,277) = 0.46, p = 0.71), and find that LLM-facilitated discussions were more information dense than unfacilitated discussions on average, where density is defined as the ratio of covered facts to the number of non-facilitator messages in the discussion ($\Delta$ = 0.067, 95\% CI [0.025, 0.109], t(138) = 3.17, p = 0.002; see Figure \ref{fig:info-sharing}B); together, these findings indicate that the increase in fact coverage is not solely a mechanical consequence of longer discussions. Moreover, the information covered in LLM-facilitated groups may be more evenly distributed among the three candidate cities than in groups with no facilitation, as suggested by the decrease in the average Gini coefficient of covered facts for each city ($\Delta$ = -0.05, 95\% CI [-0.091, -0.009], t(138) = -2.42, p = 0.017; Figure \ref{fig:info-sharing}C), hinting that LLM facilitation may encourage groups to more holistically evaluate the decision at hand.

By the volume and length of their messages, human facilitators appeared to adopt a more conversational style than their LLM counterparts. Human facilitators communicated more frequently, sending 9 messages to their group on average, while the LLM facilitator sent an average of 7 messages (by design, as detailed in Appendix \ref{app:facilitator-implementation}; t(138) = 2.74, p = 0.007), yet the average human facilitator message was much shorter with a length of 9 words, compared to the average LLM facilitator message containing 79 words. However, by evaluating the \textit{content} of the facilitator messages, we find that one potential mechanism driving the increase in information sharing is that of direct engagement with group members to participate in the discussion (e.g., ``Red, we haven't heard from you...''). On average, 25\% of a human facilitator's messages in a discussion directly addressed a group member, while LLM facilitators did so in 68\% of the messages they shared within a single discussion; the LLM facilitator directly addressed a group member at least once in every group it interacted with, yet only 77\% of human facilitators ever did so.

\begin{figure}
    \centering
\includegraphics[width=1\linewidth]{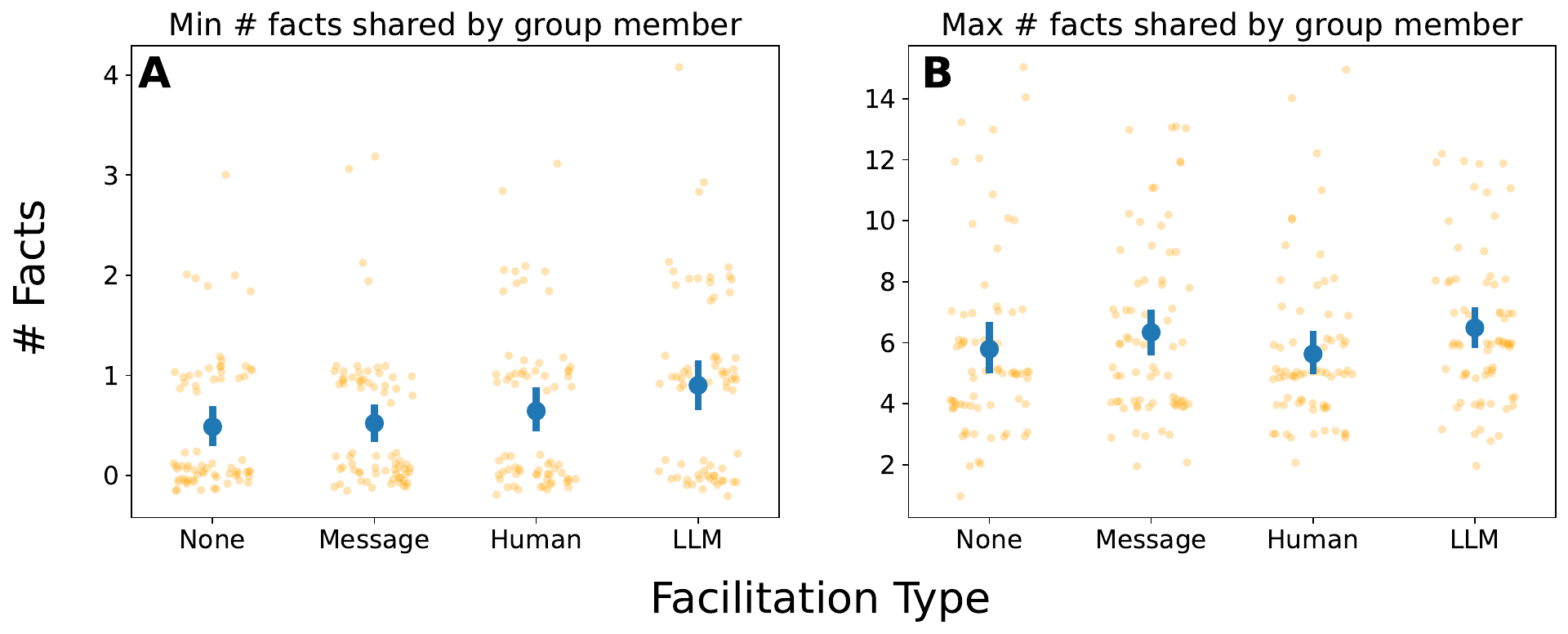}
    \caption[LLM facilitation increased the minimum individual contribution to fact aggregation]{\textbf{LLM facilitation increased the minimum individual contribution to fact aggregation.} (A) and (B) show the minimum and maximum number of distinct facts surfaced by any individual group member, respectively. Error bars indicate 95\% confidence intervals, and may be smaller than the marker. Scatter markers each represent a single group. Fact counts are discrete, but vertical jitter is added for visibility.}
    \label{fig:min-max-facts-shared}
    \Description{LLM facilitation increases the minimum number of facts contributed by any individual group member but does not meaningfully change the maximum contribution, suggesting broader participation under LLM facilitation rather than increased participation by already highly active members.}
\end{figure}

Consistent with these differences in facilitator behavior, LLM facilitation increased the minimum number of new facts (i.e., not previously covered in the group's discussion) shared by any member of the group in comparison to no facilitation ($\Delta$ = 0.41, 95\% CI [0.145, 0.684], t(138) = 3.04, p = 0.003), but not the maximum number of new facts shared by an individual (as illustrated in Figures \ref{fig:min-max-facts-shared}A and \ref{fig:min-max-facts-shared}B, respectively). Such an increase can arise through two mechanisms: an intensive margin effect, in which the least active contributors share more facts, and an extensive margin effect, in which otherwise non-participating members (those contributing no new facts) contribute at least one fact. When restricting to groups in which all members contribute at least one fact, we do not observe differences between treatments in the minimum number of facts shared (F(3,277) = 1.37, p = 0.25), suggesting limited differences in minimum engagement among already-contributing members. However, while only 39\%, 44\%, and 47\% of groups achieved contribution from all members in the ``None'', ``Message'', and ``Human'' conditions respectively, 61\% of LLM-facilitated groups had all members contribute at least one new fact ($\chi^2(3,N=281)$ = 8.10, p = 0.044), providing suggestive evidence that gains in information sharing may arise from the extensive margin.  

\subsection{Information aggregation did not translate into UWLM-consistent decisions}\label{ms:final-decisions}
Defining a group's collective choice as the option chosen by a majority, Eldoron (the optimal choice) was chosen by 31\% of unfacilitated groups, 21\% of groups presented with the one-time message, 30\% of groups facilitated by a human, and 23\% of groups facilitated by an LLM; however, these proportions are not statistically distinguishable from one another ($\chi^2(3,N=281)$ = 2.85, p = 0.42). While unaffected by the facilitative treatments, these proportions of optimal outcomes are in line with those reported for the hidden profile task upon which ours is based \cite{AmesUnknown-vy}. If groups strictly followed UWLM to make their decision, Eldoron would have been chosen by 60\% of unfacilitated groups, 48\% of groups presented with the one-time message, 47\% of groups facilitated by a human, and 63\% of groups facilitated by an LLM, yet these differences between treatments are also indistinguishable ($\chi^2(3,N=281)$ = 5.6, p = 0.13). Similarly, the proportion of groups who would have chosen Eldoron under UWLM but did not actually do so does not vary significantly across treatments ($\chi^2(3,N=281)$ = 7.52, p = 0.057). 

Beyond the stringent assumptions of strict UWLM decision-making, differential weighting of information across options may partially explain deviations from UWLM decisions. As estimated by a multinomial logit model with Myloria (the ``bait'') as the baseline option, the odds of choosing Eldoron (the optimal option) increased as its score increased ($\beta = 0.395$, 95\% CI [0.245, 0.545], p $<$ 0.001) and decreased with increasing Myloria score ($\beta = -0.734$, 95\% CI [-0.919, -0.548], p $<$ 0.001), while the score of Cragnio did not influence this contrast ($\beta = 0.054$, 95\% CI [-0.170, 0.279], p = 0.64). Similarly, the odds of choosing Cragnio increased as its score increased ($\beta = 0.417$, 95\% CI [0.136, 0.698], p = 0.004) and decreased with increasing Myloria score ($\beta = -0.649$, 95\% CI [-0.843, -0.455], p $<$ 0.001), while the score of Eldoron did not influence this contrast ($\beta = -0.043$, 95\% CI [-0.236, 0.150], p = 0.66). In both comparisons, the magnitude of the effect of Myloria's score exceeded that of the alternative city by 86\% (vs. Eldoron) and 56\% (vs. Cragnio), suggesting that group decisions are more strongly associated with information about Myloria than with information about the other options.

\begin{figure}[hbt!]
    \centering
\includegraphics[width=1\linewidth]{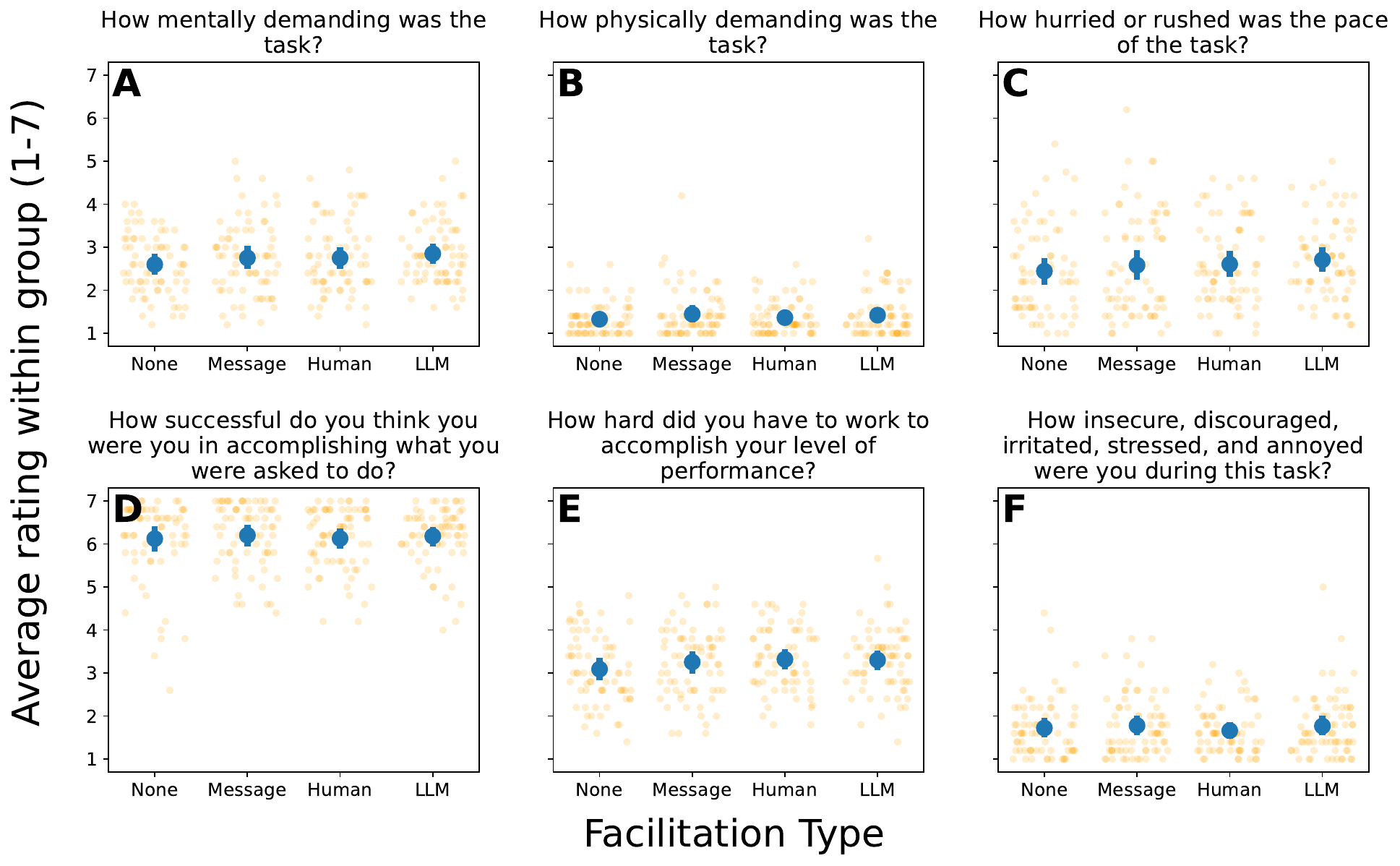}
    \caption[NASA Task Load Index (TLX)]{\textbf{NASA Task Load Index (TLX).} The TLX measures perceptions of a workload along six dimensions: mental demand (A), physical demand (B), temporal demand (C), performance (D), effort (E), and frustration (F). Participants individually answered each question on a discrete scale of 1 (``very little'') to 7 (``very much''). Responses were averaged within each group, and each scatter marker represents a single group. Error bars indicate 95\% confidence intervals, and may be smaller than the marker.}
    \label{fig:tlx-evals}
    \Description{Participants report similar workload across all facilitation conditions on all six NASA TLX dimensions, with no meaningful differences in perceived task demand.}
\end{figure}

\begin{figure}[p]
    \centering
\includegraphics[width=1\linewidth]{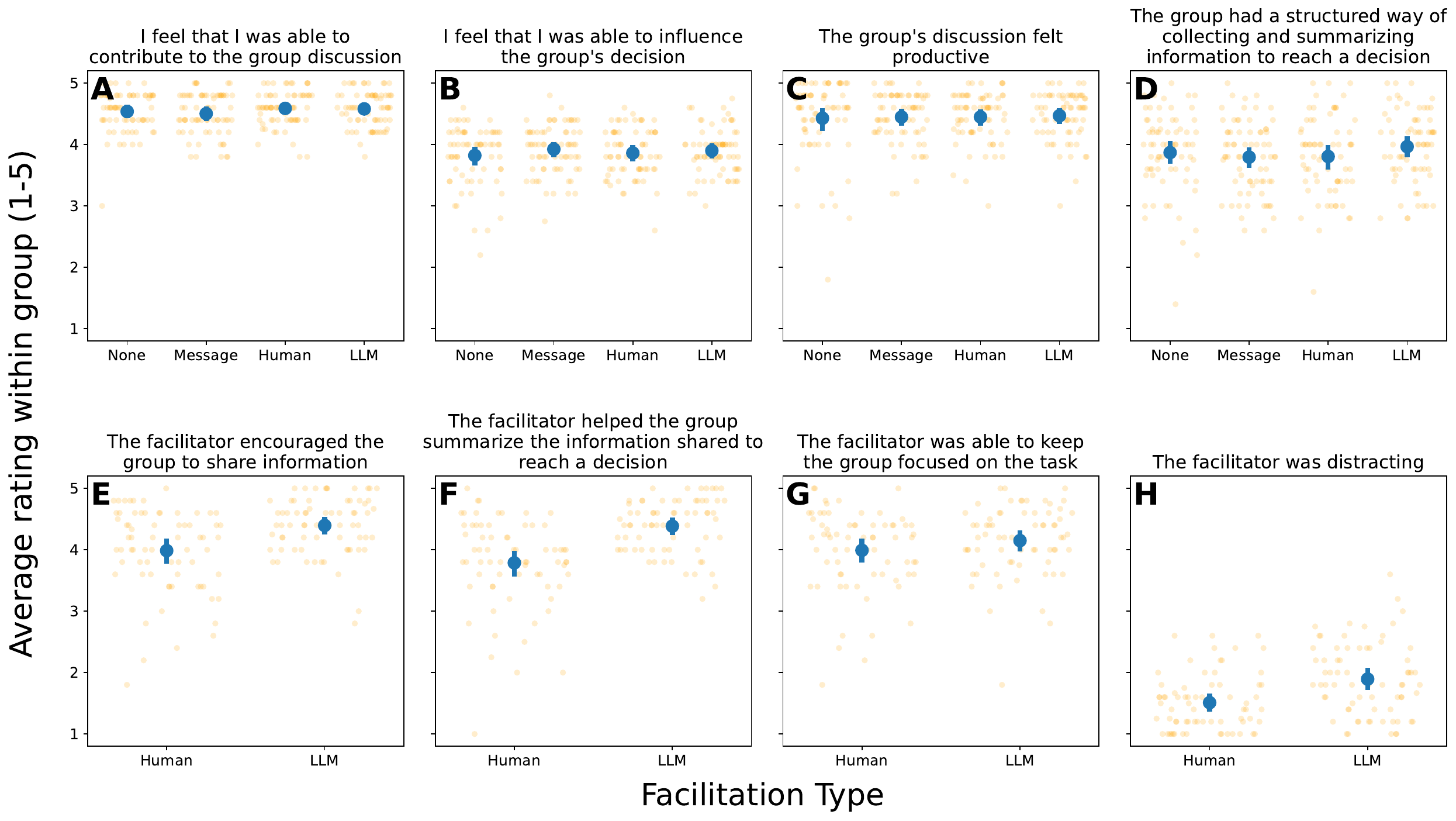}
\caption[Subjective evaluations of group dynamics and facilitator effectiveness]{\textbf{Subjective evaluations of group dynamics and facilitator effectiveness.} Participants were individually asked to evaluate their group (A-D) and facilitator (E-H) by rating their agreement with each statement on a discrete scale of 1 (``strongly disagree'') to 5 (``strongly agree''). Responses were averaged within each group, and each scatter marker represents a single group. Error bars indicate 95\% confidence intervals, and may be smaller than the marker.}
    \label{fig:subjective-evals}
    \Description{Participants rate group dynamics similarly across conditions, while LLM facilitators receive higher ratings than human facilitators for eliciting information and summarizing discussion but are also rated as slightly more distracting.}
\end{figure}

\begin{figure}[p]
    \centering
\includegraphics[width=0.6\linewidth]{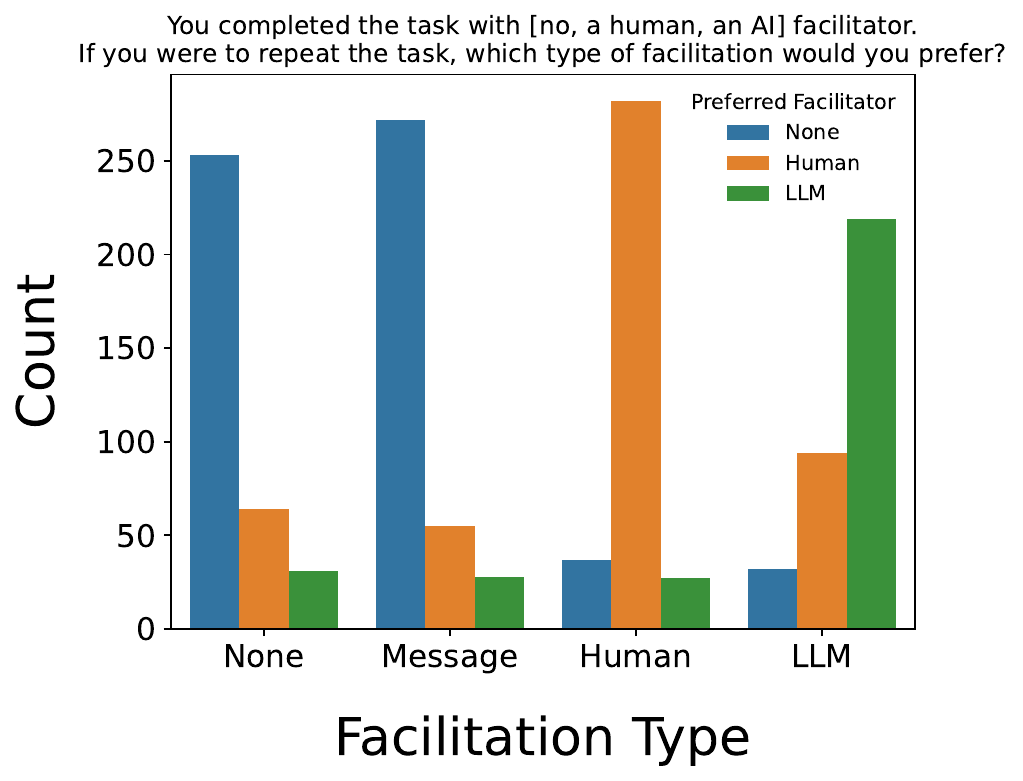}
    \caption[Participant preferences for future facilitation]{\textbf{Participant preferences for future facilitation.} Participants individually reported their preferred facilitation type if they were to hypothetically repeat the task, grouped in the figure according to their initial facilitation experience (None, Message, Human, or LLM). 
    }
    \label{fig:fac-pref}
    \Description{When asked about hypothetically repeating the task under various forms of facilitation, participants generally prefer the facilitation style they experienced during the study.}
\end{figure}

\subsection{Perceptions of the task, group, and facilitator}
As measured by the NASA TLX, group-level mean responses (averaging across members within each group) indicate that sentiment towards the task within groups was generally positive and indistinguishable across experimental conditions, with participants reporting generally low levels of mental, physical, and temporal demand, as well as effort and irritation (Figure \ref{fig:tlx-evals}A-C, E-F respectively), and high levels of perceived success in the task assigned to them (Figure \ref{fig:tlx-evals}D). Similarly, groups reported high levels of perceived personal contribution, individual influence over the decision, productivity, and structure (Figure \ref{fig:subjective-evals}A-D, respectively), with no statistically meaningful differences between conditions. Group cohesion---the mean proportion of group members who said they would want to repeat the task with the same group, as opposed to a random group---ranged from 77\% in the ``Message'' condition to 80\% in the ``Human'' condition, with insignificant differences between conditions (F(3,277) = 0.21, p = 0.89). Taken together, these results suggest that participants experienced the task and group interaction similarly across conditions, despite differences in facilitation.

In both the ``Human'' and ``LLM'' conditions, facilitators were rated positively for their ability to elicit information, summarize the discussion, and keep the group on task (Figure \ref{fig:subjective-evals}E-G, respectively). The mean group rating of LLM facilitators was higher for both elicitation ($\Delta$ = 0.41, 95\% CI [0.22, 0.60], t(138) = 4.18, p $<$ 0.001) and summarization ($\Delta$ = 0.59, 95\% CI [0.39, 0.79], t(138) = 5.87, p $<$ 0.001), which was echoed in free-text feedback: 102 participants (across 57 of 70 LLM-facilitated groups) commented positively on elicitation and/or summarization. For example, one participant noted that the facilitator was ``good for summarizing the data and prompting participation...'', while another described it as helping the group ``stay on goal and make sure we were prgoressing [sic] in an organized manner.'' 

Both facilitators were rated as generally non-distracting (Figure \ref{fig:subjective-evals}H), but the LLM was rated as more distracting ($\Delta$ = 0.38, 95\% CI [0.21, 0.56], t(138) = 4.29, p $<$ 0.001), with participants describing recurring issues such as repetitive interventions, disrupted flow, and a sense of being rushed. Direct call-outs were particularly divisive. Some participants found them effective for increasing participation (e.g., ``did a good job making people who didn't talk much try to talk more,'' and ``it reminded me to start sharing more''), while others experienced them as disruptive (``a little annoying and disruptive to the flow of conversation'') or socially uncomfortable (``it made me feel a bit pressured… like it singled me out''). Overall, participants tended to prefer the facilitation style (or lack thereof) they experienced (Figure \ref{fig:fac-pref}), suggesting that preferences may be shaped by direct experience. 

%% file: contents/discussion.tex
\section{Discussion}
\subsection{What value did LLM facilitation add to the deliberative process?} 

\subsubsection{Information aggregation}
LLM facilitation increased the volume and breadth of information shared within groups, without extending the length of discussions, indicating a shift toward more information-dense deliberation. Importantly, these gains appear to arise, at least partially, from the extensive margin: rather than amplifying already active contributors, the LLM increased the minimum level of participation within groups, reducing instances of non-contribution. This distinction is critical in hidden profile tasks, where uniquely held information may reside with a single participant, and failure to elicit that information can prevent the group from accessing the full set of relevant facts.

By increasing the likelihood of contribution and, consequently, the likelihood that privately held information enters the discussion, the LLM facilitator can improve the reliability of information elicitation. At the same time, this intervention targets availability rather than use: while more information is surfaced, how that information is interpreted, weighted, and integrated into a final decision remains unchanged. As such, LLM facilitation addresses one necessary condition for effective collective decision making---information aggregation---without resolving downstream challenges of information integration.

\subsubsection{Decision-making}
Despite increased information sharing, LLM-facilitated groups were no more likely to select the UWLM-optimal option, suggesting a disconnect between information availability and use. In one sense, this is not surprising: the model hinges on the strong assumptions that information is perceived as intended, weighted consistently, aggregated linearly, and tabulated exactly to identify the score-maximizing option. However, this disconnect is not necessarily problematic, as the goal of facilitation is not to induce strictly UWLM-consistent decisions. Rather, UWLM serves as a diagnostic benchmark for assessing whether failures arise from insufficient information aggregation or from how that information is ultimately used. In this light, prior work summarized by Lu et al.\ \cite{Lu2012-hh}, reviewing 24 empirical HPT studies, shows that groups are substantially more likely to select the correct option when profiles are manifest (i.e., all members begin with access to the full set of information), suggesting that UWLM-consistent behavior may be plausible for this task structure.

In contrast to the manifest condition, hidden profile settings introduce asymmetries in initial information that can generate early impressions, which in turn shape conversational dynamics, attention, and the interpretation of subsequently shared information. As a result, even when an intervention increases information coverage, groups may not fully recover the benefits of a manifest profile, as newly surfaced information is filtered through these initial biases. This perspective is consistent with prior work arguing that HPT performance is driven more by information \textit{use} than \textit{coverage} \cite{Xiao2016-dp}; as Dennis puts it, ``you can lead a group to information, but you can't make it think'' \cite{Dennis1996-ge}. Notably, while the LLM increased information coverage, this shift remains modest relative to the full information set, leaving groups in a partially hidden regime where early asymmetries can continue to shape how information is interpreted and used.

In our setting, although the information weighting we observe is directionally consistent with UWLM decision-making (higher normative scores increase the odds of selection), we find that information supporting the ``bait'' option (Myloria) is more strongly associated with selection than information supporting the other options. While our design does not identify the mechanism underlying this asymmetry, this difference may reflect the individual-level preference bias commonly documented in the HPT paradigm \cite{Faulmuller2012-ls, Mojzisch2010-ol, Kelly1999-wk, Schulz-Hardt2016-og}. Because each participant initially sees Myloria as the most appealing option in their individual information set, they may remain inclined to focus on Myloria in their \textit{use} of the covered facts, thereby influencing group-level decision making. 

\subsubsection{Subjective experience}
Despite meaningfully altering the structure of group discussions, LLM facilitation did not negatively impact participants' experience of the task or their group. Across measures of workload and group dynamics, we observe no meaningful differences between conditions, suggesting that the gains in information sharing were not achieved at the expense of increased cognitive burden or reduced group cohesion.

Relative to the human facilitator, the LLM outperformed in eliciting and summarizing information, despite relying on a relatively simple, domain-agnostic setup. However, participant feedback also highlights trade-offs in how these interventions were experienced, particularly around disruption and direct targeting. Taken together, these findings empirically underscore the need for facilitation systems that address not only what to say, but when and to whom--an emphasis reflected in emerging frameworks for group-AI intervention design \cite{Mao2024-tb, Do2022-am, Papachristou2025-ex}.

\subsection{Designing more effective LLM facilitators}\label{ms:discussion-llm-design}
\subsubsection{Process intervention beyond information aggregation}
Our findings suggest that while our instance of LLM facilitation improves the availability of information, it may not meaningfully alter how that information is interpreted, weighted, and integrated into a final decision. This gap is, in part, a consequence of our design choices: the LLM intervened at fixed intervals with group-level messages, operating only over the shared transcript and general instructions to elicit and organize information. As a result, the facilitator primarily influenced what information entered the discussion, without shaping how that information was subsequently processed.

One implication is that more targeted process interventions could influence how information is perceived as the discussion unfolds. For example, rather than continuously aggregating information, facilitation could segment deliberation into cycles of aggregation and reflection, enabling groups to periodically synthesize and evaluate what has been shared before introducing additional information. More broadly, the appropriate structure of information aggregation is likely to be context-dependent. While the LLM in our study encouraged more balanced discussion across options, such holistic coverage may be suboptimal in time-constrained settings where dominated alternatives should be quickly eliminated. This suggests value in allowing users to specify or adapt the facilitation process to the demands of the task. Process interventions may also be applied at the decision-making stage. Introducing structured procedures such as staged evaluation, ranking, or variants of the Delphi method, could support more systematic comparison of alternatives and reduce the influence of early impressions on final decisions.

Importantly, we deliberately constrained the LLM to a process-oriented role, avoiding direct recommendations to preserve the group's sense of ownership over the decision. Future systems that more directly shape the decision-making process (eg., by pointing to critical trade-offs or contributing judgment) may help guide more effective information use, but introduce new risks around bias and over-reliance. The ``Deliberative AI'' system described by Ma et al.\ \cite{Ma2025-aj} integrates both process and content interventions in a single-user system, and suggests a promising scaffold to be extended to multi-user settings. 

\subsubsection{Managing coordination}
The subjective evaluations of the LLM facilitator point to a trade-off between coordination and attention management in group-AI interactive settings. Although participants appreciated the LLM's role in aggregating and organizing information, they also perceived it as more distracting than a human facilitator. Feedback describing excessive or disruptive interjections, and a sense of being rushed, suggests that facilitation competes for participant attention within a shared communication channel. 

In addition to the content of an LLM intervention, its form may shape this trade-off. In our study, LLM facilitator messages were substantially longer than those written by human facilitators and were delivered repeatedly throughout a fixed 10-minute discussion. This format introduces both attentional and temporal pressure. Longer facilitator messages occupy more of the visible chat window, potentially crowding out participant messages and interrupting conversational flow. They also require more time to process, potentially contributing to the sense of feeling rushed and reducing the time group members have to share, discuss, and interpret information to make a decision. 

Design strategies such as constraining message length, adjusting intervention frequency, or shifting toward more conversational, incremental interactions may reduce this burden. However, such changes may also affect the effectiveness of facilitation, introducing a tension between informativeness and intrusiveness that warrants further study. A complementary approach is to diversify the channels through which LLM facilitators engage participants. Rather than addressing the entire group simultaneously, facilitative systems could also interact with participants individually. This would enable the facilitator to privately deliver feedback (as studied by Do et al. \cite{Do2022-am}), discuss concerns or motivations, or provide tailored prompts aligned with each participant's level of engagement, potentially reducing distraction in the shared discussion space while preserving the benefits of targeted process intervention.

\subsubsection{Managing emotional tension}
Unlike human facilitators, LLMs may more readily directly address individual group members. Because these call-outs are visible to the entire group, they can increase social pressure to participate, potentially reducing social loafing \cite{Jackson1985-fl}. However, participant feedback suggests that these interventions can also induce discomfort. Prior work by Houtti et al.\ \cite{Houtti2025-eh} suggests that when interacting with AI facilitators, group members may respond to such discomfort by rationalizing away or dismissing critical feedback, potentially undermining the effectiveness of the intervention.

This suggests another dynamic trade-off: while direct interventions may boost immediate participation, they may also erode receptivity to future interventions if they generate sufficient discomfort. As a result, the effectiveness of facilitation may depend not only on whether interventions occur, but on how they shape participants' willingness to engage with subsequent guidance, necessitating careful calibration of when and how individuals are targeted.

\subsection{Limitations}
The effects of LLM facilitation on information sharing, decision-making, and subjective experience are shaped by several design choices in our experimental setting. Beyond the implementation of the LLM facilitator and its interface, as discussed in Section \ref{ms:discussion-llm-design}, contextual factors such as time pressure \cite{Kelly1999-wk, Kelly2004-dj}, group size \cite{Cruz1997-dy}, and communication modality (chat, video, audio, face-to-face) \cite{Shirani2006-rk, Lam2000-pr, Kerr2009-ws} can significantly alter the attentional landscape of the task, in turn influencing the efficacy of LLM facilitation.

Holding the design of the LLM facilitator and task context constant, the characteristics of the hidden profile task itself may further shape these effects. For example, an HPT's difficulty \cite{Schulz-Hardt2006-os, Brodbeck2002-in} and total information load \cite{Stasser1987-cy} may moderate both a facilitator's ability to elicit information and the decision-making value of any additional information that is surfaced. More fundamentally, the HPT in our study is a collaborative task in a neutral domain; effectively managing divergent viewpoints and incentives in contentious \cite{Manata2019-ix} or competitive \cite{Wittenbaum2004-oe, Toma2009-qm} settings may require more sophisticated facilitation strategies than those considered here. 

In sum, the range of factors influencing LLM-facilitated group interactions necessitate a systematic and integrative approach to future research \cite{Almaatouq2022-ml}. By carefully examining these dimensions jointly, the field can advance toward more robust and generalizable insights into effective group-AI facilitation.

\subsection{Open data and open-source software to support group-AI interaction research}
Progress in understanding group-AI interaction is constrained by limited access to shared datasets and experimental infrastructure. While paradigms such as the HPT provide a canonical setting for studying how groups aggregate distributed information, results are often difficult to compare or build upon due to differences in implementation and the lack of accessible, process-level data. Advancing toward a cumulative science of group-AI interaction requires both richly instrumented datasets and reusable experimental platforms.

To this end, we release our dataset capturing 14,343 messages shared by 1,475 participants across 281 groups, providing a rich resource for researchers studying facilitative technologies and group decision-making. Alongside the transcripts, we provide all LLM requests and responses, along with associated metadata and LLM-generated rationales, enabling researchers to conveniently perform post-hoc analyses or simulate alternative facilitation prompts.

In addition to the dataset, we introduce the Group-AI Interaction Laboratory (GRAIL), our open-source experimental platform based on the Empirica framework \cite{Almaatouq2021-cs}, joining a growing library of group-AI deliberation research tools such as the Deliberate Lab platform \cite{Qian2025-eu}. GRAIL is designed to address key practical challenges in conducting group-based behavioral experiments, including coordinating synchronous participation, managing real-time interaction, and integrating LLMs into shared communication environments. By providing a flexible interface for specifying task structure, participant-specific stimuli, and LLM intervention logic, GRAIL allows researchers to prototype and deploy new experimental designs without re-implementing the underlying system.

%% file: contents/conclusion.tex
\section{Conclusion}
Our study demonstrates the practical capacity of LLMs to \textit{actively} support group decision making discussions by directly engaging group members and prompting participation as a discussion evolves, as well as effectively summarizing rapid, information-dense communication. By raising the minimum level of individual task engagement  within groups, the LLM facilitator increased both the breadth of information at a group's disposal, and the likelihood that \textit{all} group members contributed information to the decision---an outcome that is critical in settings with asymmetric information, where a single member's private knowledge may be pivotal. 

Decision-making, whether performed individually or collaboratively, is a complex, multi-stage process. Although the increased information sharing observed did not translate directly into better final decisions in our experiment, this outcome highlights the inherent complexity and context dependence of group decision-making. It also underscores the potential value of integrating computational support that focuses distinctly on each stage of the deliberative process, from initial information gathering through to evaluation and consensus formation. 

By encouraging broader participation within groups, the LLM facilitator helps groups consider a wider range of perspectives, particularly from group members who might otherwise remain passive. In practical scenarios analogous to hidden profile tasks---such as hiring committees---this form of facilitation may play an important role in mitigating biases through promoting conversational equity. Moreover, continued development of computational facilitation technologies could expand access to these organizational techniques, making them available to communities unable to afford professional facilitation services.  

Our work builds upon an extensive body of research within the CSCW and broader HCI communities studying the effectiveness of group decision support systems, which have historically incorporated a wide range of computational methods. The novel affordances of LLMs present an opportunity to increase the fidelity and adaptivity of these systems, and it is our hope that the robust empirical evidence of this potential that we have presented here inspires future work in this direction, and that our open-source experimental platform enables researchers to systematically explore and extend these potential benefits.

%% file: contents/acknowledgments.tex
\section{Data, code, and institutional review}
Experimental data and analysis code are available on the study's Open Science Framework repository (\href{https://doi.org/10.17605/OSF.IO/ERVNB
}{https://doi.org/10.17605/OSF.IO/ERVNB}), and the GRAIL platform is available through its GitHub repository (\href{https://github.com/microsoft/group_ai_lab}{https://github.com/microsoft/group\_ai\_lab}). The study was approved by the Microsoft Research Institutional Review Board (MSR IRB; Record \#10883), and all participants provided explicit consent to participate. 

\begin{acks}
We are grateful to Daniel Ames for his helpful insight into the hidden profile task paradigm. 
\end{acks}

%% file: contents/appendix.tex
\appendix
\section{Recruitment details and materials} \label{app:recruitment}
\begin{figure}[h]
    \centering
\includegraphics[width=1\linewidth]{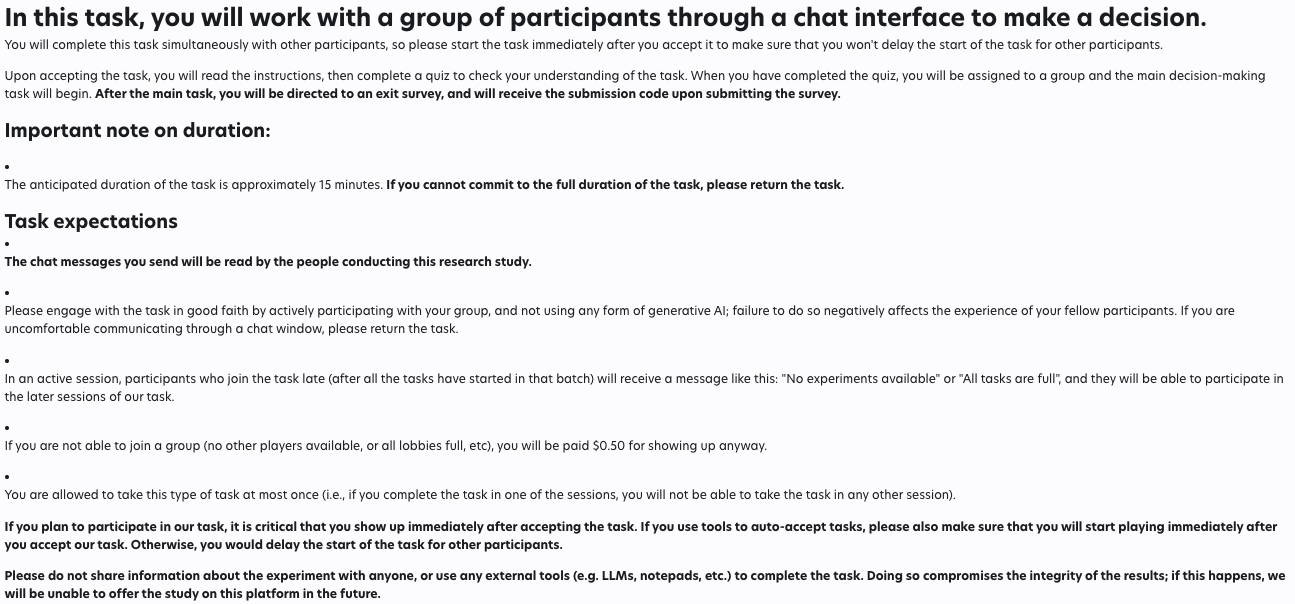}
    \caption[Prolific recruitment text]{\textbf{Prolific recruitment text}}
    \label{fig:recruitment-text}
    \Description{A screenshot of the task instructions as presented on Prolific, outlining task expectations and requirements.}
\end{figure}

A total of 1,475 participants were recruited from Prolific through the platform's standard sample, filtered to those from the United States with self-declared fluency in English, more than 50 previous submissions to Prolific, and a submission acceptance rate greater than 95\%; the exact recruitment text is shown in Figure \ref{fig:recruitment-text}. Participants who completed all steps of the study, including the exit survey, were paid a flat rate of \$3.75 USD. Participants were only allowed to participate once. To facilitate simultaneous recruitment of hundreds of participants within minutes, the tasks on Prolific were configured to disable the default ``rate limiting'' that Prolific imposes on the distribution of tasks to participants.

\section{Pre-registration details}\label{app:prereg}
The study design and main analysis of the number of facts shared within groups were pre-registered (\href{https://aspredicted.org/23df-gtys.pdf}{AsPredicted \#192061}). We note two slight deviations:
\begin{itemize}
    \item We pre-registered that we would retain data from groups where every participant sent at least 1 message to the group, the group collectively sent at least 10 messages, and every participant submitted the task completion code. In a slight deviation from the last criterion, we checked that participants' submissions were accepted on Prolific by the research team, either by submitting the correct completion code or manually by the researchers, as some participants faced technical issues where, for example, the submission code did not appear at the end, or they were unable to submit the code on Prolific. 

    \item As pre-registered, data was collected until the pre-registered count of 70 valid groups per experimental condition was met, with the exception of the ``Message'' treatment, for which 71 valid groups were collected as a consequence of ensuring that groups could be assigned to any treatment with equal probability. 
\end{itemize}

\section{Screenshots and implementation details for treatment conditions}\label{app:interface-screenshots}
\subsection{``None'' and ``Message'' treatments}
\begin{figure}[h]
    \centering
\includegraphics[width=1\linewidth]{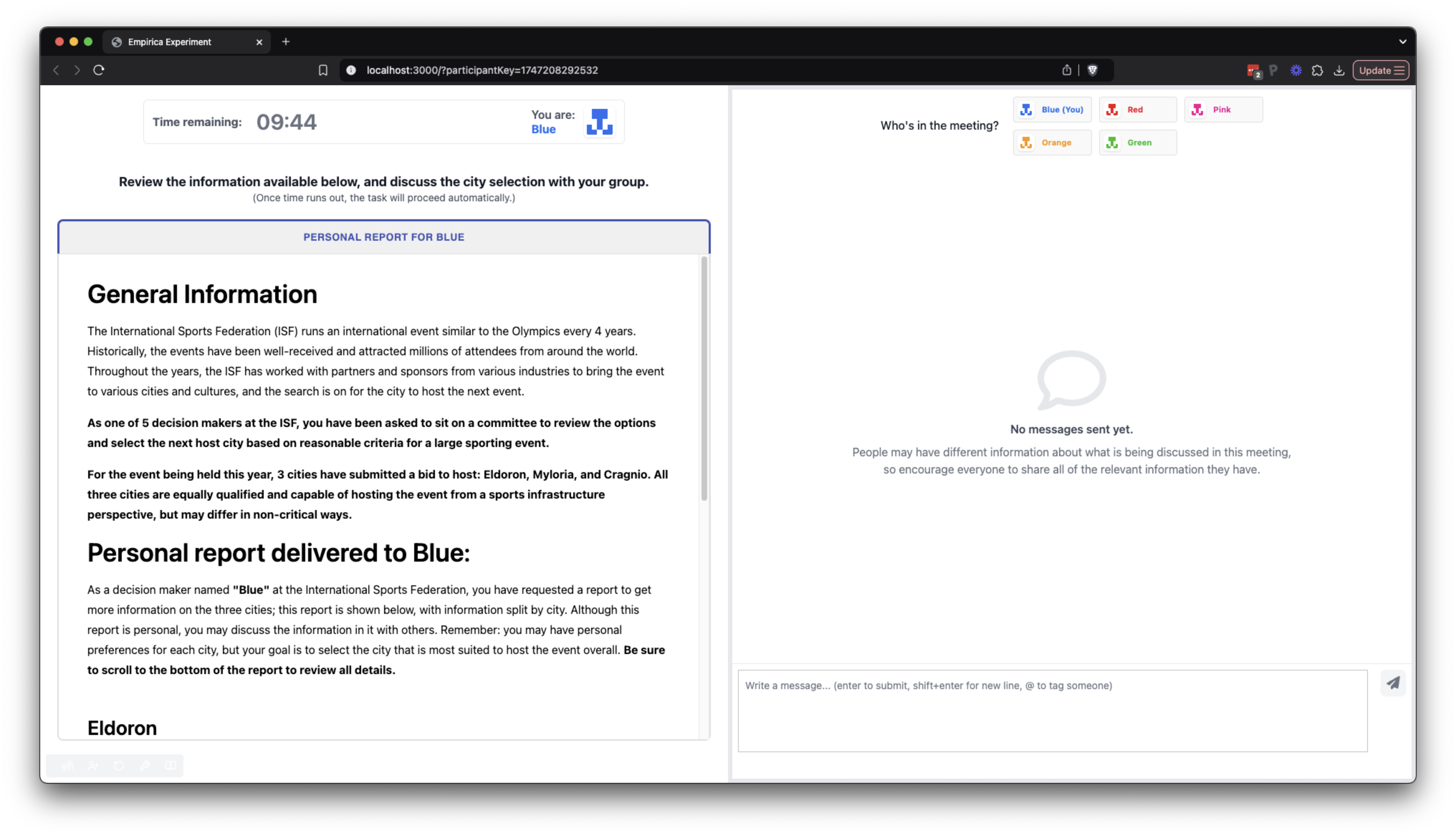}
    \caption[Initial information sharing message]{\textbf{Initial information sharing message}}
    \label{fig:cue1}
\end{figure}

\begin{figure}[h]
    \centering
\includegraphics[width=1\linewidth]{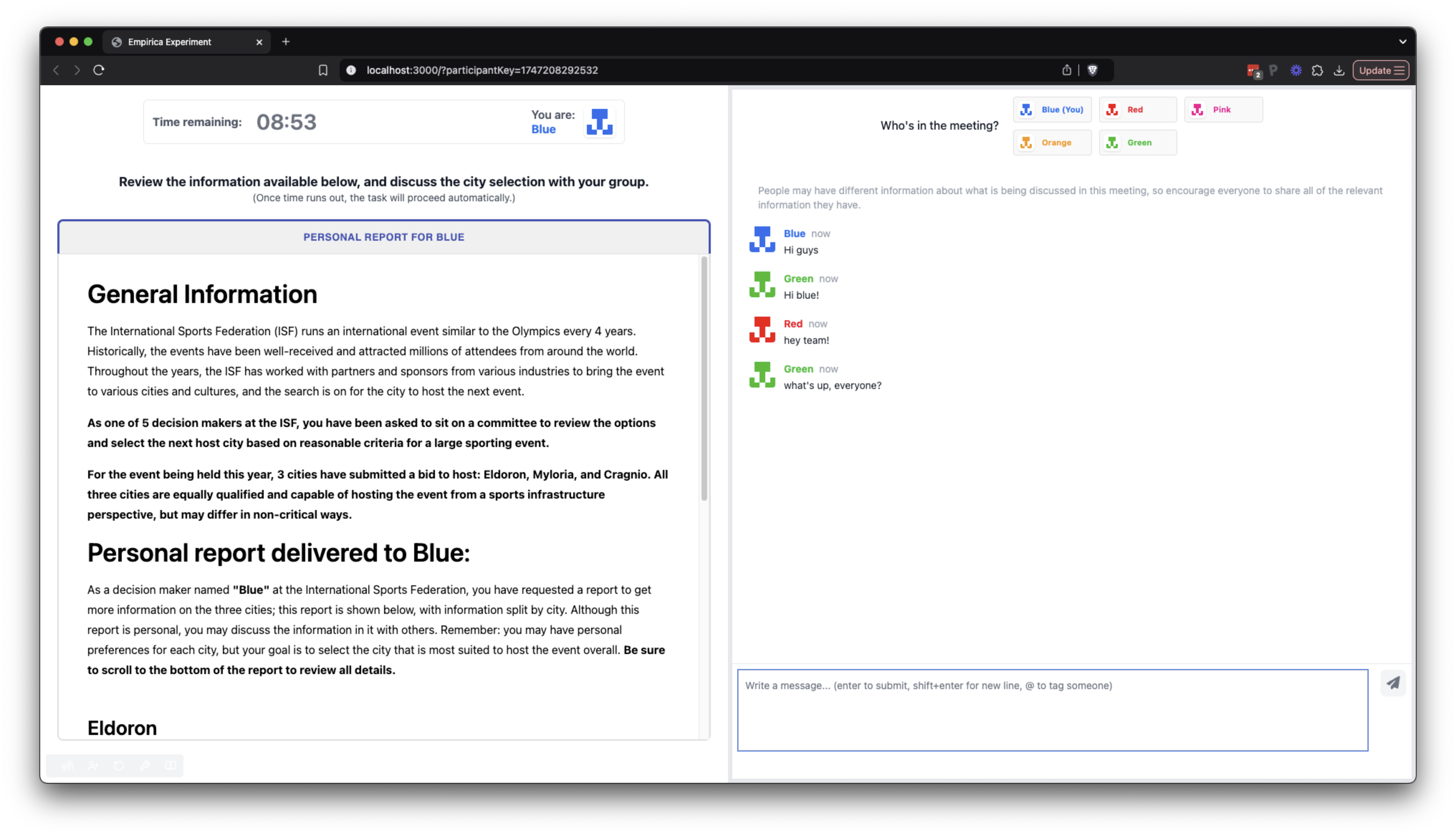}
    \caption[Information sharing message moving to the top of the conversation]{\textbf{Information sharing message moving to the top of the conversation}}
    \label{fig:cue2}
\end{figure}

In both the ``None'' and ``Message'' treatments, a facilitator is not added to the group. The difference between the two conditions arises in the initial layout of the chat window; in the ``Message'' condition, the empty chat window shows a reminder to participants to share information (Figure \ref{fig:cue1}). Once a group in the ``Message'' condition starts communication, the reminder moves to the top of the screen, before gradually being pushed out of view as the conversation expands (Figure \ref{fig:cue2}). 


\subsection{The ``Human'' facilitator condition}
\begin{figure}[h!]
    \centering
\includegraphics[width=1\linewidth]{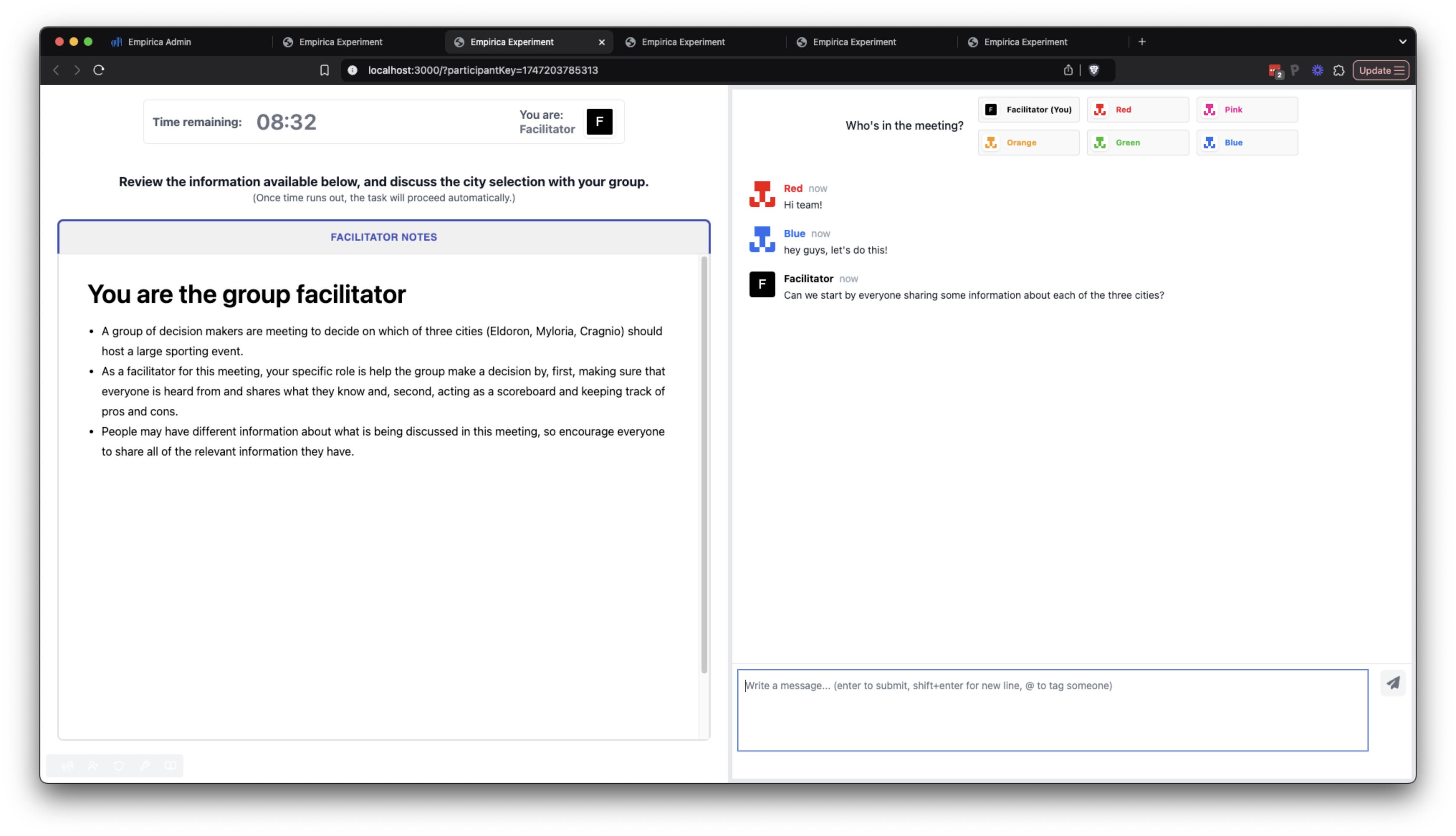}
    \caption[The human facilitator's interface]{\textbf{The human facilitator's interface}}
    \label{fig:fac_view}
\end{figure}

When a sixth participant was added to the group to act as facilitator, they saw the same chat window as other participants, but rather than receiving an information set, they were provided with the same set of instructions provided to the LLM facilitator, as shown in Figure \ref{fig:fac_view}. 

\section{Facilitator LLM Prompting and Prompt Formatting}\label{app:facilitator-implementation}
To implement the LLM facilitation condition, we configured the experimental platform to automatically query GPT-4o (gpt-4o-2024-08-06) every 90 seconds during the 10-minute decision-making task (i.e., at the 1.5, 3, 4.5, 6, 7.5, and 9 minute marks, for a total of 6 queries over the duration of the task). Additionally, as they were informed in the task introduction, groups could also choose to manually query the LLM facilitator by explicitly tagging the facilitator in conversation (by typing ``@Facilitator''). However, queries to the LLM could also fail, leading to slight variation in the number of messages that LLM-facilitated groups would receive from the LLM. 

Automatic queries were only triggered if the group had sent at least one message, and the last message sent to the group was not sent by the LLM facilitator; otherwise, the query was skipped. By contrast, manual queries were always processed. If a query---automated or manual---failed (e.g., due to API failure), this was logged, and the group would not receive a message from the LLM for that query. 

Across 70 groups in the ``LLM'' condition, with an expected total of 420 scheduled queries, only 26 LLM queries were skipped or failed, with 48 groups experiencing 6 LLM interventions (as intended), 18 groups experiencing 5 interventions, and 4 groups experiencing 4 interventions. Of the 26 scheduled LLM interventions that did not occur, 15 were skipped because the last message in the discussion had been sent by the LLM, 6 were skipped because the group had not begun discussing yet, and 5 were due to actual API failure. Out of 114 direct queries made by groups to the LLM, only 1 request had an API failure. No data were excluded as a consequence of scheduled query skipping or API failure.  

\subsection{LLM Facilitator Prompt Structure}
Each query to the LLM facilitator included the following contents in the order presented: 
\begin{itemize}
    \item \textbf{System prompt:} The full text of the system prompt is included below, and configured as a ``system'' message following the OpenAI API convention.
    \item \textbf{Meeting attendees:} ``Meeting attendees: [list of connected group member names at time of query]'', configured as a ``system'' message. 
    \item \textbf{Time Elapsed:} ``TIME ELAPSED: X minutes, Y seconds'', configured as a ``system'' message. 
    \item \textbf{Time Remaining:} ``TIME REMAINING: X minutes, Y seconds'', configured as a ``system'' message. 
    \item \textbf{Transcript at time of query:} The complete group transcript from the beginning of the task to the time of the query, where each message is presented in a new line and formatted as ``[X minutes ago] SENDER\_NAME: MESSAGE\_CONTENT''. The transcript is configured as a ``user'' message for the API.
\end{itemize}

\subsection{LLM Facilitator System Prompt}
The LLM facilitator's system prompt is as follows: 
\lstset{
basicstyle=\small\ttfamily,
columns=flexible,
breaklines=true,
breakindent=0pt,
}
\begin{lstlisting}
A group of decision makers are meeting to decide on which of three cities (Eldoron, Myloria, Cragnio) should host a large sporting event. 
As a facilitator for this meeting, your specific role is help the group make a decision by, first, making sure that everyone is heard from and shares what they know and, second, acting as a scoreboard and keeping track of pros and cons. 
People may have different information about what is being discussed in this meeting, so encourage everyone to share all of the relevant information they have.
You will periodically receive the transcript of the group's conversation so far (as well as any previous messages you've sent the group), and you will be able to provide your input.
Messages you have previously sent to the group will be included in the transcript you receive, and will be shown as sent by the "Facilitator". 
You're given the time remaining for the group to make a decision, and each message in the transcript has a timestamp. 
When given the transcript, you should respond with a JSON containing the following: 
    
MESSAGE: Include the text of your message here -- DO NOT USE MARKDOWN, but you can add newlines for formatting.
RATIONALE: Explain why you chose the message you did.
    
Remember the following: 
When intervening, you should aim to be as concise as possible while still providing all the necessary guidance.
You can tag a specific person in your message by using the "@" symbol followed by their name in square brackets (e.g. "@[NAME]"). If you do not use the square brackets, the tag will not work.
ONLY RESPOND IN JSON FORMAT, DO NOT RESPOND IN PLAIN TEXT!
\end{lstlisting}

\section{Hidden Profile Task Design}\label{app:hpt-design}
The hidden profile task used in this study was a modified version of the ``Grogan Air'' team decision exercise developed by Prof.\ Daniel Ames at Columbia Business School \cite{AmesUnknown-vy}, in which groups of 5 members review information about 3 potential candidates for an executive role at a fictional airline. The structure of the individual information sets was preserved (i.e., the polarity and visibility of facts), while the content of the facts in the information set was changed to describe fictional candidate cities to host an international sporting event akin to the Olympics. To direct groups' focus to the facts at hand, participants were informed that ``all three cities are equally qualified and capable of hosting the event from a sports infrastructure perspective, but may differ in non-critical ways'' at the beginning of their individual reports. 

\begin{figure}
    \centering
\includegraphics[width=0.85\linewidth]{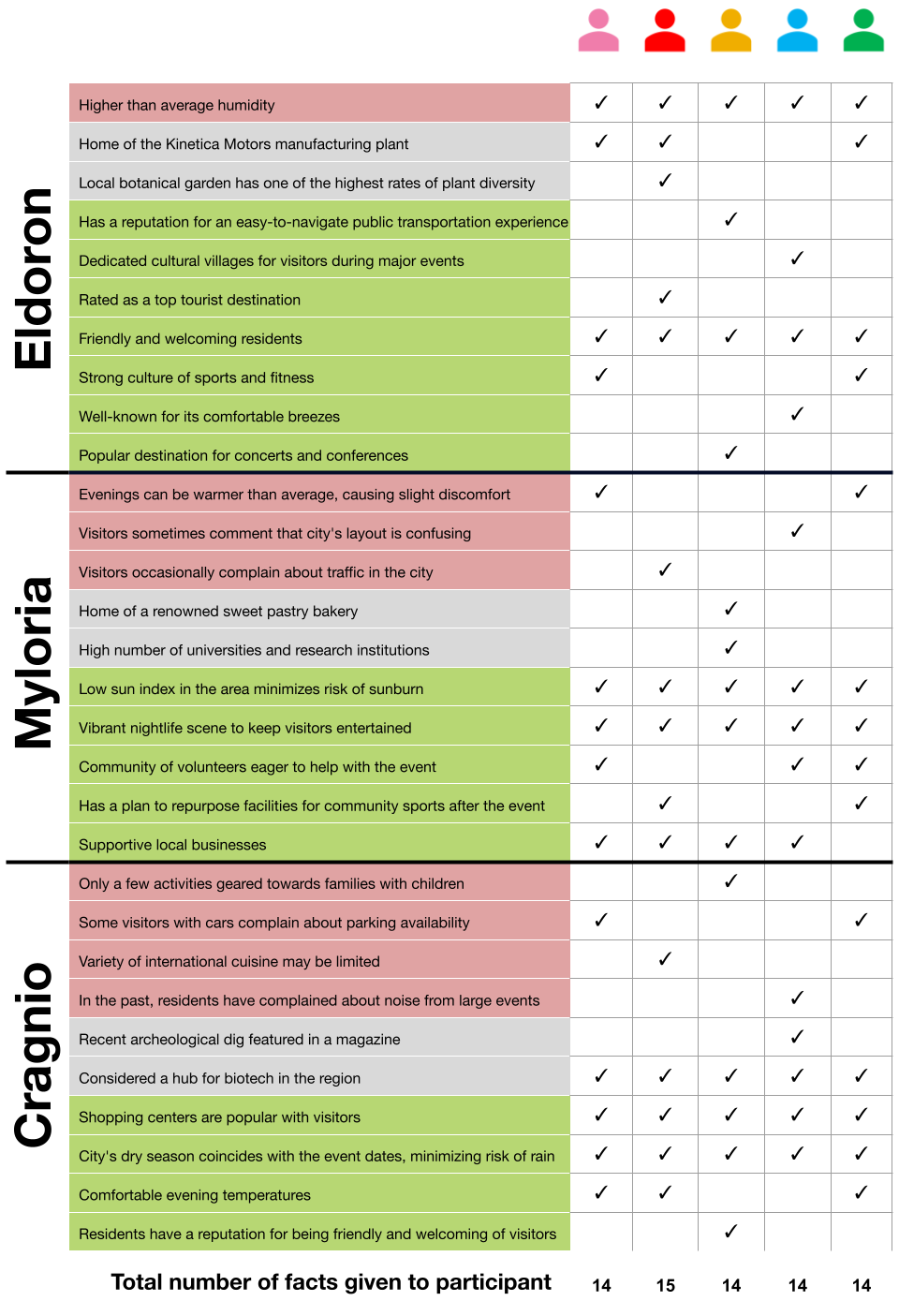}
    \caption[Hidden profile task information distribution]{\textbf{Hidden profile task information distribution:} Each of the 3 candidate cities is described by 10 facts. In turn, each fact is defined by its polarity (red for negative, grey for neutral, and green for positive) and visibility, indicated by the check marks showing which group members (Pink, Red, Orange, Blue, Green) it is visible to.}
    \label{fig:hpt-info}
\end{figure}

As in ``Grogan Air'', there is a total of 30 facts, with 10 facts describing each of the three options (Eldoron, Myloria, and Cragnio) with each individual group member's information set including a subset of 14-15 facts. Figure \ref{fig:hpt-info} shows the content of each city's fact set, and the distribution of the facts among each of the group's 5 members (referred to as Pink, Red, Orange, Blue, and Green). 

\begin{figure}
    \centering
\includegraphics[width=0.85\linewidth]{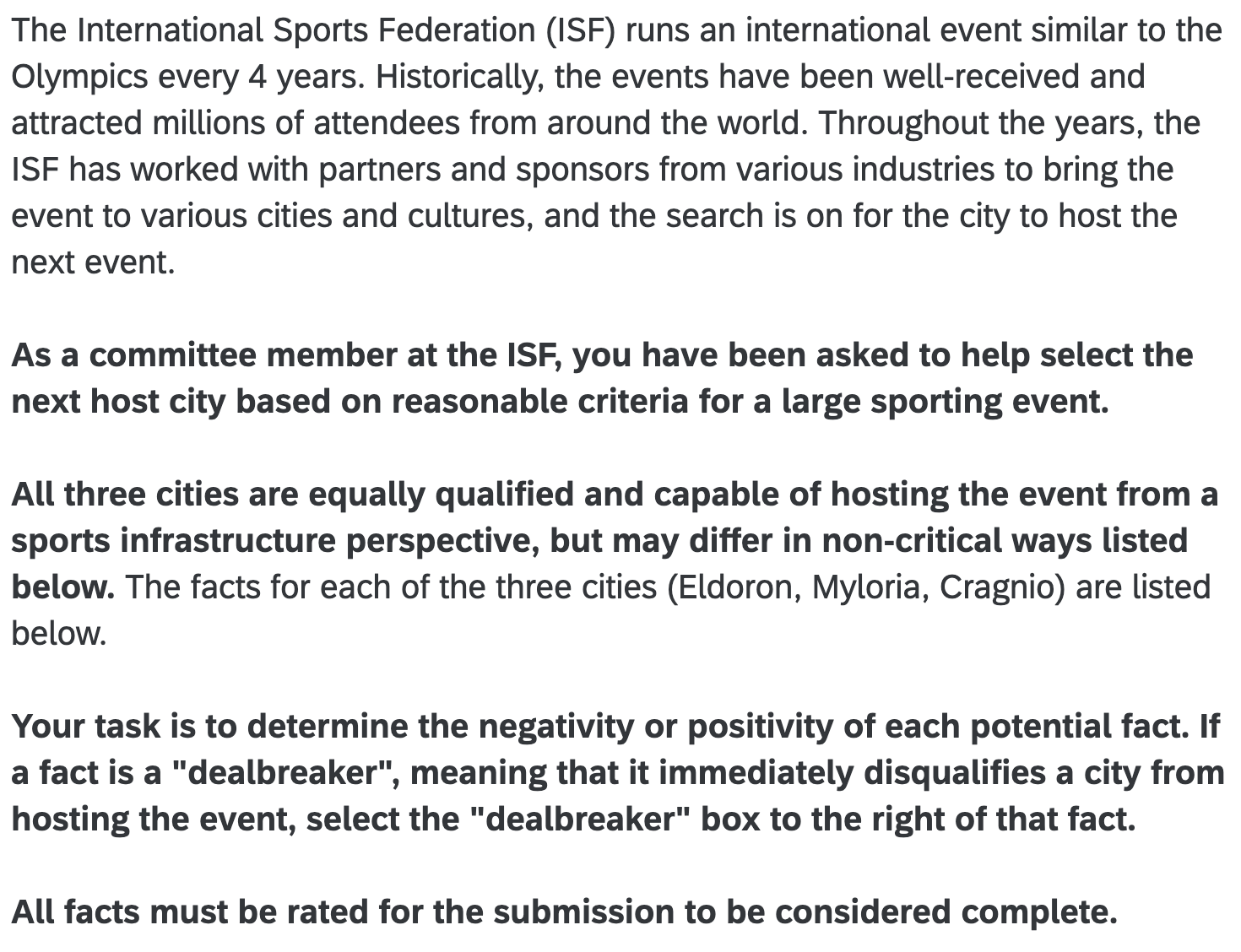}
    \caption[Fact rating survey instructions]{\textbf{Fact rating survey instructions}}
    \label{fig:fact-rating-survey-intro}
\end{figure}

Following the structure of ``Grogan Air'', we aimed to phrase facts that would be widely perceived as negatively or positively affecting a city's ability to host such an event, as well as facts that would not plausibly affect a host city's eligibility. To validate whether these facts were perceived as intended, 119 Prolific participants (who did not participate in the main experiment) were randomly assigned to one of the five information sets, presented with the same instructions provided in the main experiment, and asked to rate each fact on a Likert scale of 1 (``very negative'') to 7 (``very positive''), with 4 labeled as ``Neutral'' (see Figure \ref{fig:fact-rating-survey-intro} below for the exact instructions); the facts were presented by information set so that survey respondents would view them in the same conditions (e.g., adjacent facts) as participants from the main experiment. 

\begin{figure}
    \centering
\includegraphics[width=0.9\linewidth]{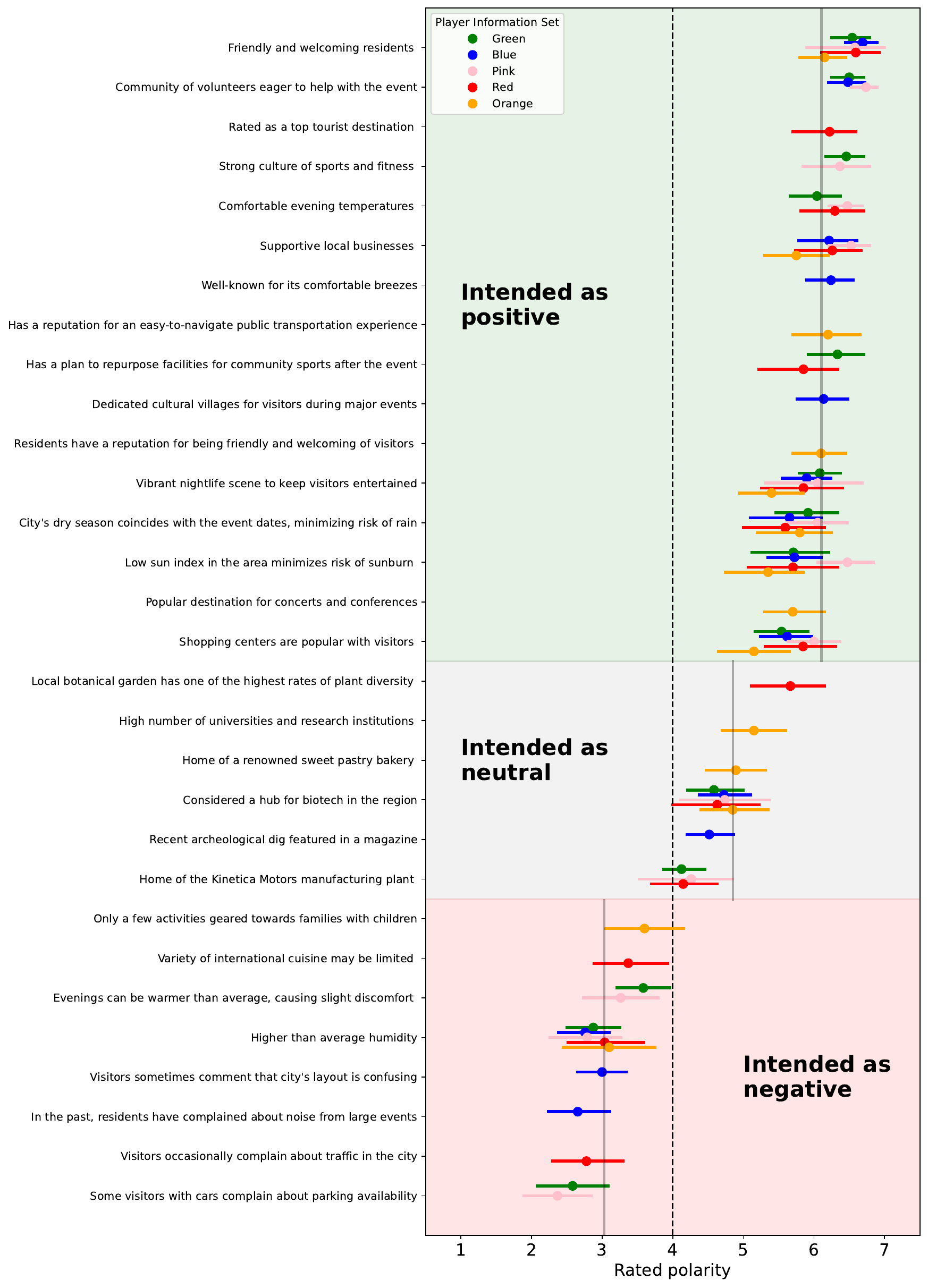}\textbf{}
    \caption[Fact polarity evaluation survey]{\textbf{Fact polarity evaluation survey:} Each survey respondent was individually shown one of the five information sets (Pink, Red, Orange, Blue, Green) and rated the facts it contained from 1 (very negative) to 7 (very positive), with 4 being described as ``neutral''. Error bars indicate 95\% CIs of the given ratings, and the grey vertical line in each cluster of positive/neutral/negative facts indicates the within-cluster average of average fact ratings across information sets.}
    \label{fig:fact-rating}
\end{figure}

As shown in Figure \ref{fig:fact-rating}, facts about the cities were generally perceived as intended, with the exception of some neutral facts leaning towards positive. Out of 30 facts, 25 were rated as ``dealbreakers'' by less than 5\% of respondents, with the most ``dealbreaking'' fact being ``In the past, residents have complained about noise from large events'', as rated by 17\% of respondents; in aggregate, these results suggest the absence of overwhelmingly influential facts.

\section{Using an auxiliary LLM to detect fact mentions in group discussion}\label{app:fact-detection}
The hidden profile task used in the study includes  a total of 30 facts. To detect when these facts are mentioned in a group's discussion, we prompt GPT-4o (gpt-4o-2024-08-06) with instructions for the fact detection task and a list of all 30 facts (as shown in Appendix \ref{app:fact-detection-prompt} below). Each group discussion is processed as a single transcript, rather than processing each message individually, to allow the LLM to detect implicit references, e.g., saying ``yeah but the parking there is pretty bad'' in response to an earlier mention of Cragnio. 

The LLM's fact detection performance was validated against a set of 5 randomly selected group transcripts manually labeled with mentioned facts by the authors; this validation set included a total of 272 messages, 177 of which included no facts, 70 of which included 1 fact, and 25 of which included more than 1 fact. The lists of facts detected by the humans and LLM annotator match exactly in 95\% of the 272 test cases; all 14 cases in which the annotators differed involved a single fact that the human annotator detected as an implicit mention, while the LLM annotator did not. 

The inclusion of facilitator messages in the annotated transcripts introduces another way to verify the consistency of annotations. Because the facilitators do not have access to an information set, their messages should not introduce new facts to the discussion, and any facts the annotator detects in a facilitator's message must be detected in a message preceding it in the discussion. Out of 1,133 messages sent by facilitators, only 2 fail this checksum. In the first case, a fact was correctly detected in a facilitator's summary but was missed in the original participant message where it first appeared. In the second case, the annotator incorrectly attributes two additional facts to a facilitator's message; while the cause is unclear, these same facts are correctly detected in the immediately following participant message, meaning the error does not affect group-level fact coverage.  

In general, these findings suggest that the LLM fact detection measure works acceptably well; any peculiarities in its performance would likely affect conversations across the different experimental treatments similarly, and thus not affect the differences we observe between treatments.  

\subsection{Fact detection system prompt}\label{app:fact-detection-prompt} The system prompt used for the auxiliary, post-hoc fact detecting LLM is as follows: 

\lstset{
basicstyle=\small\ttfamily,
columns=flexible,
breaklines=true,
breakindent=0pt,
}
\begin{lstlisting}
A group of executives met to discuss hosting an event in one of three cities: Eldoron, Myloria, and Cragnio. 
Below are pieces of information about each city; you will be given the executives' conversation, and should return lists of the pieces of information that are mentioned in each message from the conversation. 

For example, if the message was 'well, yeah, but cragnio has noise complaints and good weather', the correct response would be to return ['C6', 'C10'] for that message. 
Only include a fact if it is directly referenced. For example, a message saying 'ok, let's start by discussing eldoron' or 'i think that climate/weather is a particularly important consideration for the comfort of our visitors' does not refer to any single fact.

Each message in the conversation is presented in the format [MESSAGE ID]: SPEAKER: TEXT. 

RETURN A JSON OBJECT WHERE THE KEY IS THE MESSAGE ID, AND THE VALUE IS THE LIST OF FACT IDS MENTIONED IN THAT MESSAGE. 
IF NO FACTS ARE MENTIONED IN A MESSAGED, RETURN AN EMPTY LIST AS THE VALUE FOR THAT MESSAGE. 
EVERY MESSAGE SHOULD HAVE AN ENTRY IN THE RETURNED JSON OBJECT. 

The bank of facts is:

Cragnio
C1: Residents have a reputation for being friendly and welcoming of visitors 
C2: City's dry season coincides with the event dates, minimizing risk of rain
C3: Shopping centers are popular with visitors 
C4: Considered a hub for biotech in the region
C5: Recent archeological dig featured in a magazine
C6: In the past, residents have complained about noise from large events
C7: Variety of international cuisine may be limited 
C8: Some visitors with cars complain about parking availability
C9: Only a few activities geared towards families with children
C10: Comfortable evening temperatures 

Eldoron
E1: Higher than average humidity
E2: Well-known for its comfortable breezes
E3: Home of the Kinetica Motors manufacturing plant 
E4: Local botanical garden has one of the highest rates of plant diversity 
E5: Popular destination for concerts and conferences
E6: Has a reputation for an easy-to-navigate public transportation experience
E7: Dedicated cultural villages for visitors during major events
E8: Rated as a top tourist destination 
E9: Friendly and welcoming residents 
E10: Strong culture of sports and fitness 

Myloria
M1: Visitors sometimes comment that city's layout is confusing
M2: Supportive local businesses 
M3: Has a plan to repurpose facilities for community sports after the event
M4: Community of volunteers eager to help with the event
M5: Vibrant nightlife scene to keep visitors entertained
M6: Low sun index in the area minimizes risk of sunburn 
M7: Home of a renowned sweet pastry bakery 
M8: Visitors occasionally complain about traffic in the city
M9: Evenings can be warmer than average, causing slight discomfort 
M10: High number of universities and research institutions 
\end{lstlisting}

\section{Screenshots of the introduction and exit survey interfaces}\label{app:intro-exit}
\subsection{Introduction and attention check} Participants are introduced to the purpose of the meeting and the task to be completed, then complete a brief attention check. 
\begin{figure}[h!]
    \centering
\includegraphics[width=0.85\linewidth]{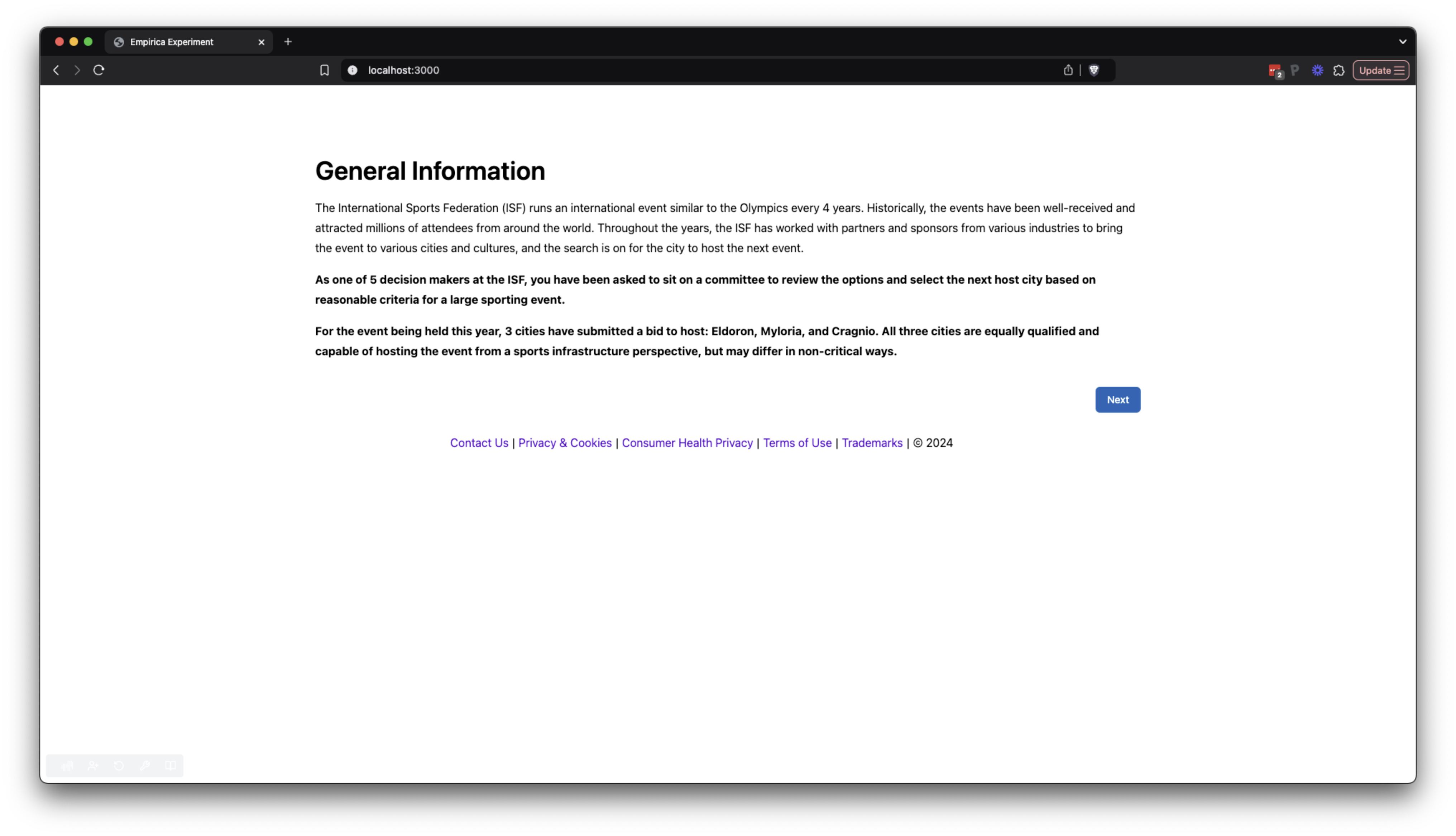}
\end{figure}
\begin{figure}[h!]
    \centering
\includegraphics[width=0.85\linewidth]{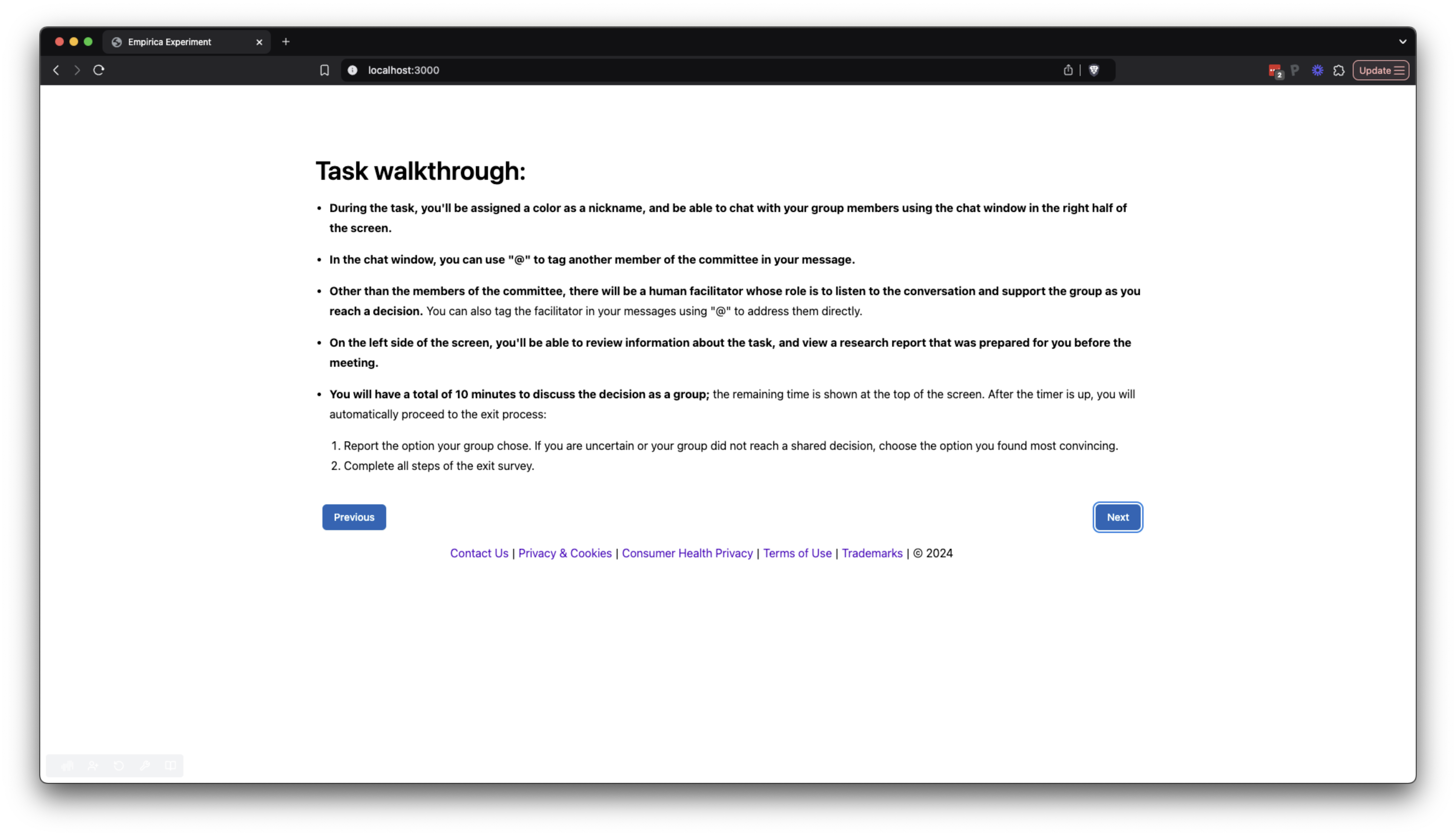}
\end{figure}
\begin{figure}[h!]
    \centering
\includegraphics[width=0.85\linewidth]{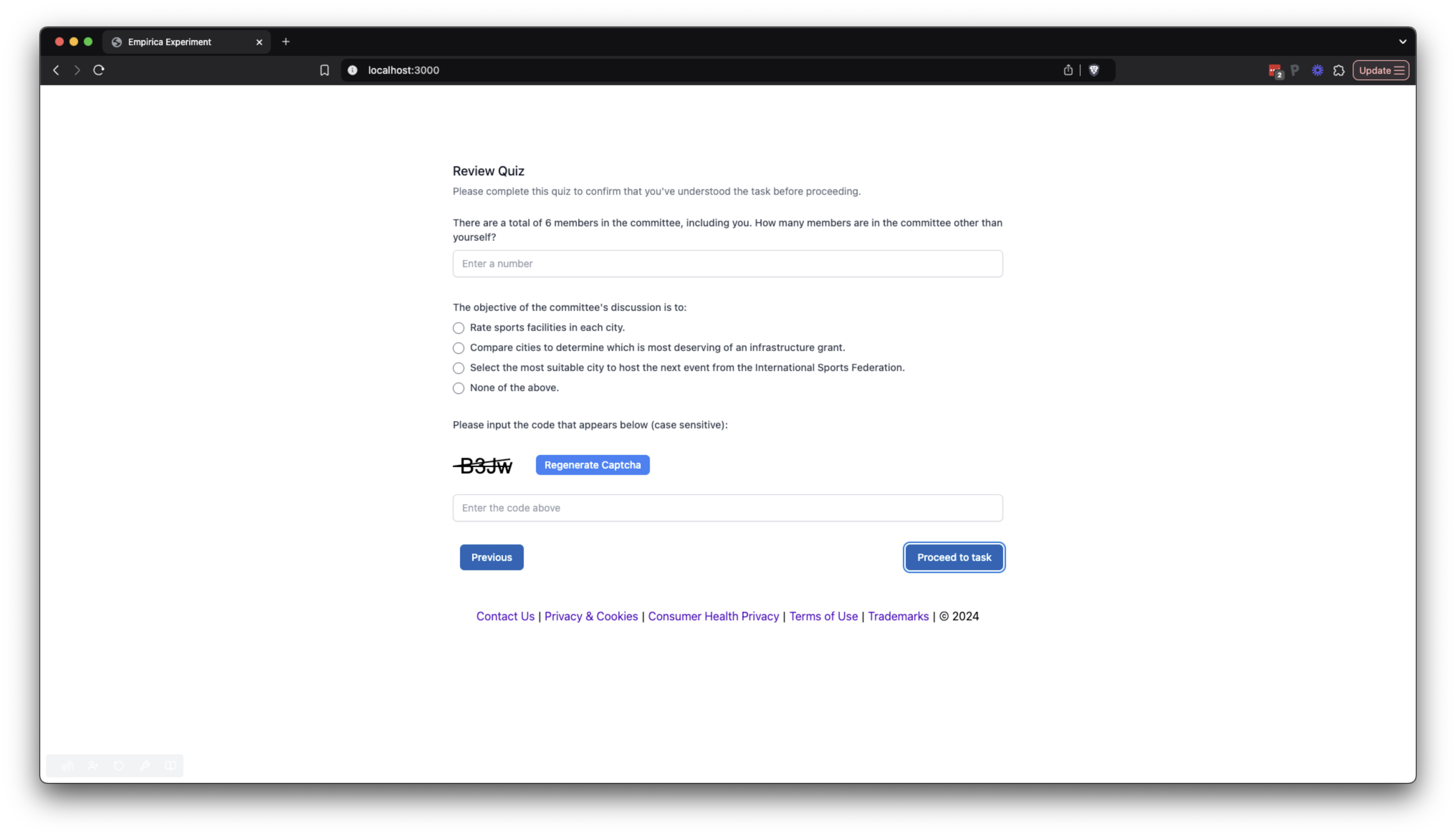}
\end{figure}

\subsection{Decision reporting and exit survey} Participants individually and asynchronously report the decision their group made, as well as their subjective evaluations of the task, their group, and their facilitator if they had one. 

\begin{figure}[h]
    \centering
    \includegraphics[width=0.85\linewidth]{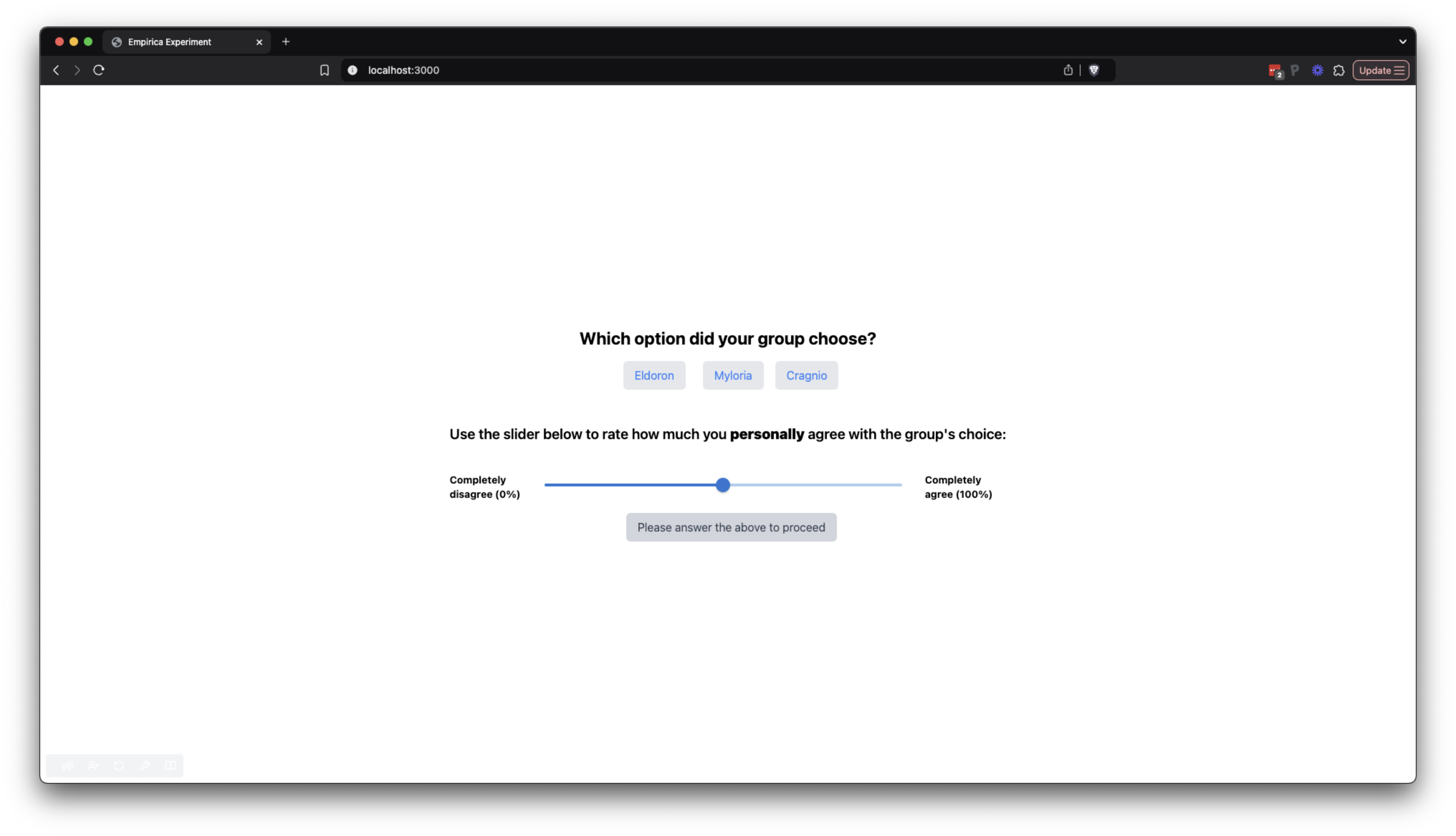}    \includegraphics[width=0.85\linewidth]{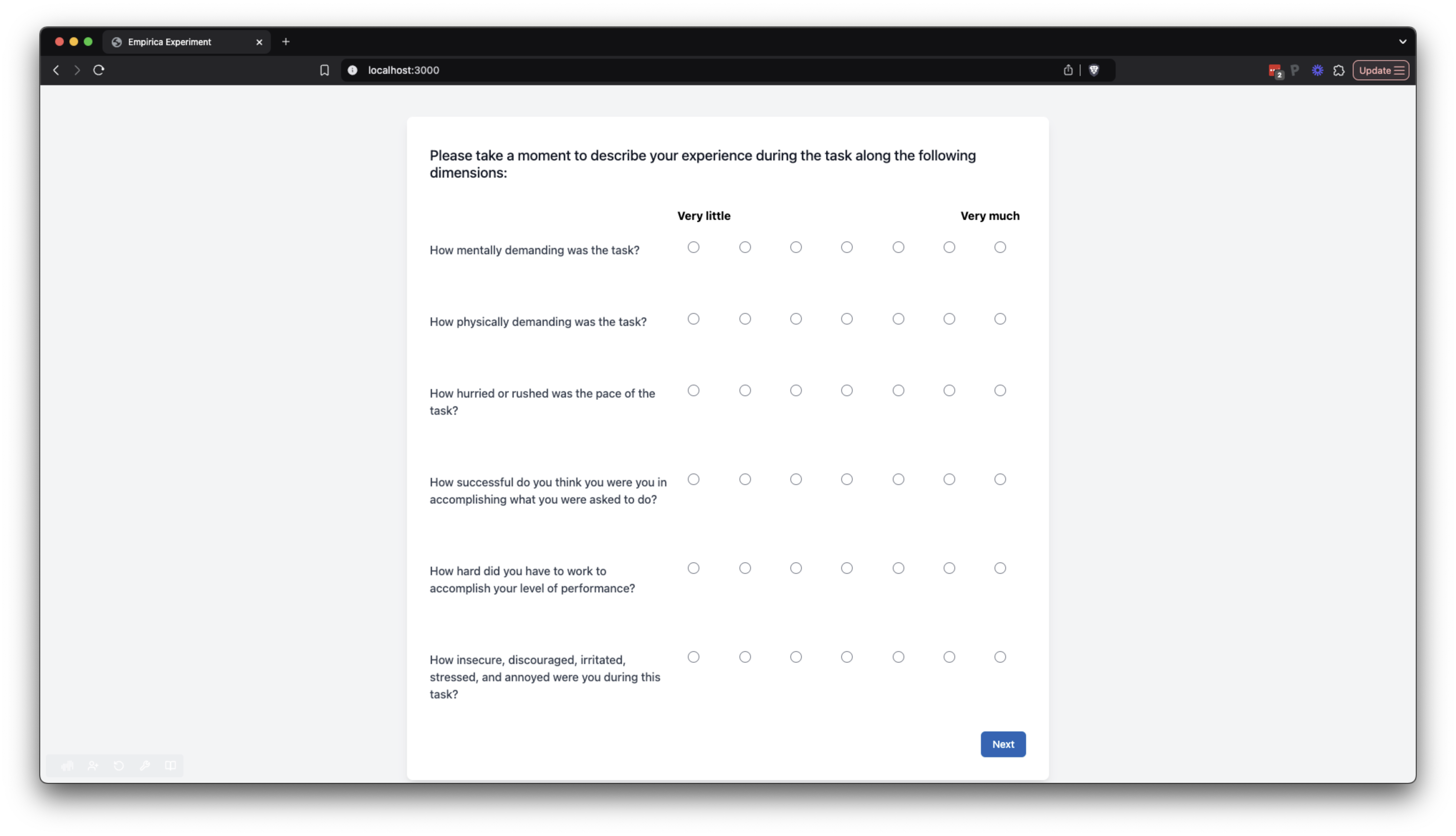}
\end{figure}

\begin{figure}[h]
    \centering
    \includegraphics[width=0.85\linewidth]{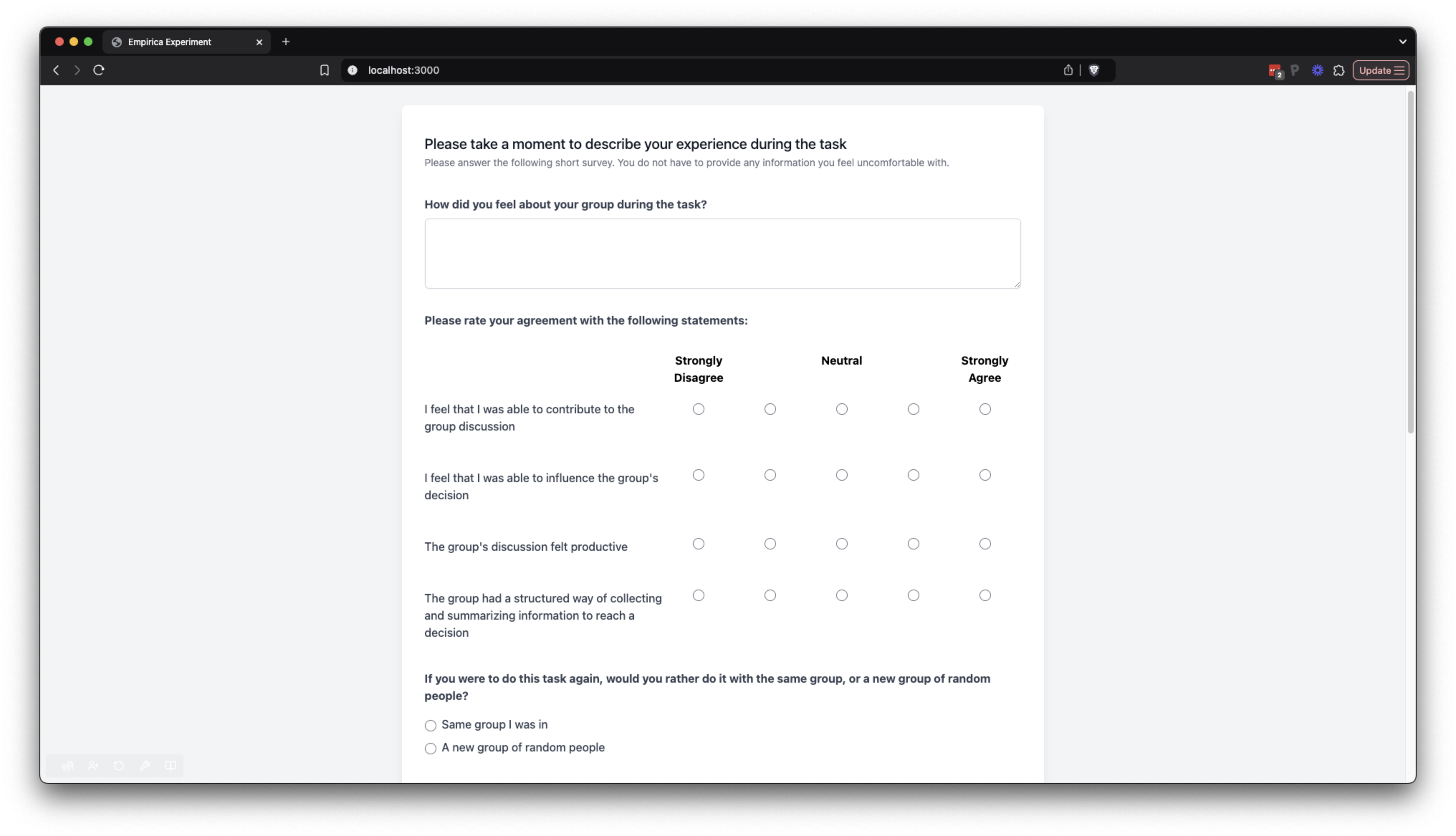}    \includegraphics[width=0.85\linewidth]{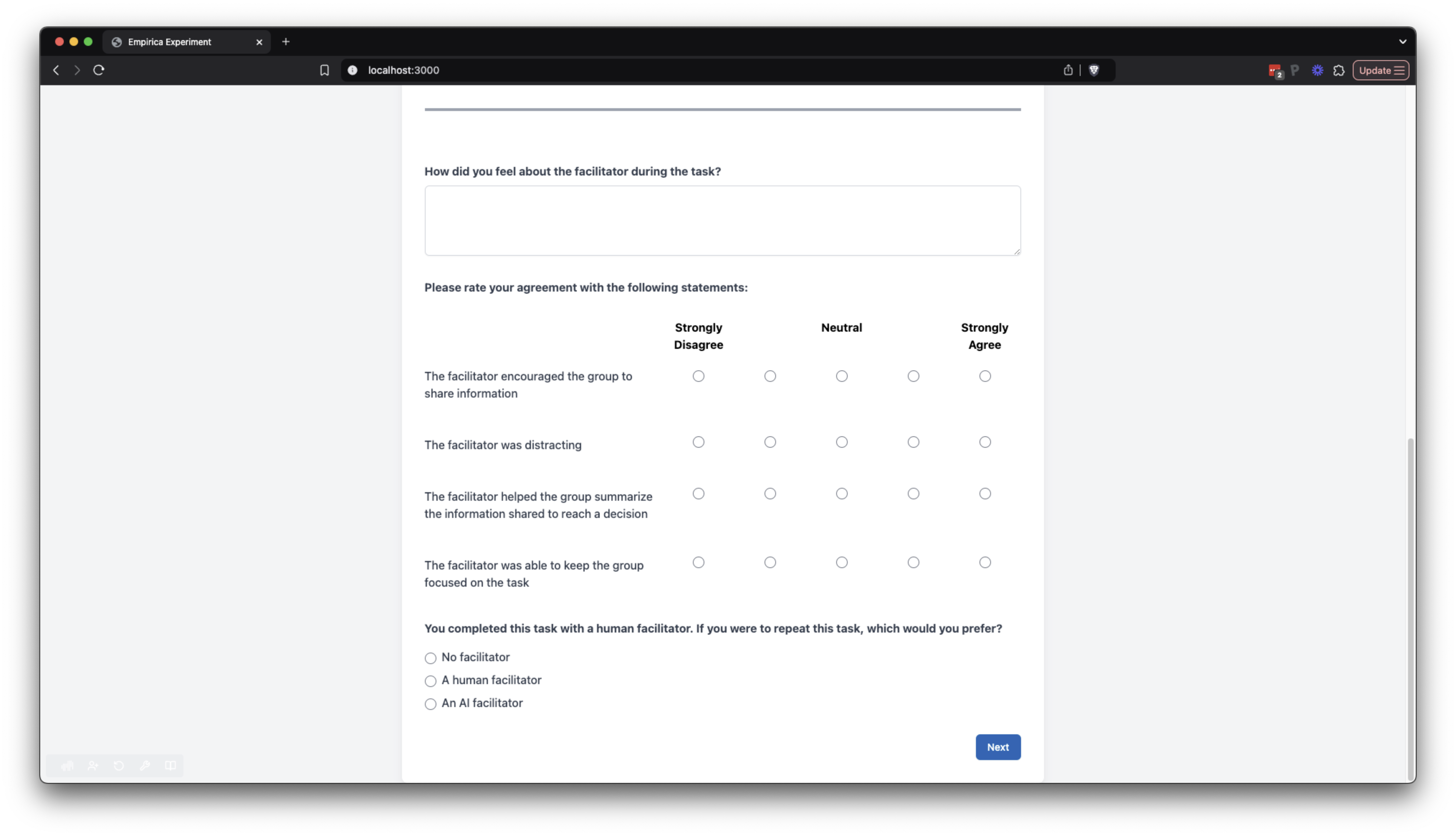}
\end{figure}

%% file: main.bib
@ARTICLE{Sohrab2015-bs,
  title     = "Exploring the Hidden-Profile Paradigm: A Literature Review and
               Analysis",
  author    = "Sohrab, Serena G and Waller, Mary J and Kaplan, Seth",
  journal   = "Small Group Research",
  publisher = "SAGE Publications Inc",
  volume    =  46,
  number    =  5,
  pages     = "489--535",
  month     =  "1~" # oct,
  year      =  2015
}

@ARTICLE{Do2022-am,
  title     = "How should the agent communicate to the group? Communication
               strategies of a conversational agent in group chat discussions",
  author    = "Do, Hyo Jin and Kong, Ha-Kyung and Lee, Jaewook and Bailey, Brian
               P",
  journal   = "Proc. ACM Hum. Comput. Interact.",
  publisher = "Association for Computing Machinery (ACM)",
  volume    =  6,
  number    = "CSCW2",
  pages     = "1--23",
  month     =  "7~" # nov,
  year      =  2022,
  language  = "en"
}

@ARTICLE{Lu2012-hh,
  title    = "Twenty-five years of hidden profiles in group decision making: a
              meta-analysis",
  author   = "Lu, Li and Yuan, Y Connie and McLeod, Poppy Lauretta",
  journal  = "Pers. Soc. Psychol. Rev.",
  volume   =  16,
  number   =  1,
  pages    = "54--75",
  month    =  feb,
  year     =  2012,
  language = "en"
}

@ARTICLE{Nicholson2021-gk,
  title     = "'{I}'ve Just Been Pretending {I} Can See This Stuff!': Group
               member voice in decision-making with a hidden profile",
  author    = "Nicholson, Dawn H and Hopthrow, Tim and de Moura, Georgina
               Randsley and Travaglino, Giovanni A",
  journal   = "Br. J. Soc. Psychol.",
  publisher = "Wiley",
  volume    =  60,
  number    =  3,
  pages     = "1096--1124",
  month     =  jul,
  year      =  2021,
  language  = "en"
}

@INCOLLECTION{Viller1991-uz,
  title     = "The Group Facilitator: A {CSCW} Perspective",
  author    = "Viller, Stephen",
  booktitle = "Proceedings of the Second European Conference on
               Computer-Supported Cooperative Work ECSCW ’91",
  publisher = "Springer Netherlands",
  address   = "Dordrecht",
  pages     = "81--95",
  year      =  1991
}

@INPROCEEDINGS{Shamekhi2018-gs,
  title     = "Face value? Exploring the effects of embodiment for a group
               facilitation agent",
  author    = "Shamekhi, Ameneh and Liao, Q Vera and Wang, Dakuo and Bellamy,
               Rachel K E and Erickson, Thomas",
  booktitle = "Proceedings of the 2018 CHI Conference on Human Factors in
               Computing Systems",
  publisher = "ACM",
  address   = "New York, NY, USA",
  month     =  "21~" # apr,
  year      =  2018
}

@ARTICLE{Christakis2008-xv,
  title    = "The collective dynamics of smoking in a large social network",
  author   = "Christakis, Nicholas A and Fowler, James H",
  journal  = "N. Engl. J. Med.",
  volume   =  358,
  number   =  21,
  pages    = "2249--2258",
  month    =  "22~" # may,
  year     =  2008,
  language = "en"
}

@ARTICLE{Larson1993-ul,
  title     = "Groups as problem-solving units: Toward a new meaning of social
               cognition",
  author    = "Larson, Jr, James R and Christensen, Caryn",
  journal   = "Br. J. Soc. Psychol.",
  publisher = "Wiley",
  volume    =  32,
  number    =  1,
  pages     = "5--30",
  month     =  mar,
  year      =  1993
}

@ARTICLE{Almaatouq2021-nf,
  title    = "Task complexity moderates group synergy",
  author   = "Almaatouq, Abdullah and Alsobay, Mohammed and Yin, Ming and Watts,
              Duncan J",
  journal  = "Proc. Natl. Acad. Sci. U. S. A.",
  volume   =  118,
  number   =  36,
  month    =  "7~" # sep,
  year     =  2021,
  language = "en"
}

@BOOK{Larson2010-mo,
  title     = "In search of synergy in small group performance",
  author    = "Larson, Jr, James R",
  publisher = "Psychology Press",
  address   = "New York, NY, US",
  year      =  2010
}

@ARTICLE{Jackson1985-fl,
  title   = "Social loafing on difficult tasks: Working collectively can improve
             performance",
  author  = "Jackson, Jeffrey M and Williams, Kipling D",
  journal = "J. Pers. Soc. Psychol.",
  volume  =  49,
  number  =  4,
  pages   = "937--942",
  month   =  oct,
  year    =  1985
}

@ARTICLE{Manata2019-ix,
  title     = "Assessing the Effects of Partisan Bias at the Group Level of
               Analysis: A Hidden Profile Experiment",
  author    = "Manata, Brian and Boster, Franklin J and Wittenbaum, Gwen M and
               Bergan, Daniel E",
  journal   = "American Politics Research",
  publisher = "SAGE Publications Inc",
  volume    =  47,
  number    =  6,
  pages     = "1283--1302",
  month     =  "1~" # nov,
  year      =  2019
}

@ARTICLE{Almaatouq2021-cs,
  title    = "Empirica: a virtual lab for high-throughput macro-level
              experiments",
  author   = "Almaatouq, Abdullah and Becker, Joshua and Houghton, James P and
              Paton, Nicolas and Watts, Duncan J and Whiting, Mark E",
  journal  = "Behav. Res. Methods",
  volume   =  53,
  number   =  5,
  pages    = "2158--2171",
  month    =  oct,
  year     =  2021,
  language = "en"
}

@ARTICLE{Almaatouq2022-ml,
  title    = "Beyond playing 20 questions with nature: Integrative experiment
              design in the social and behavioral sciences",
  author   = "Almaatouq, Abdullah and Griffiths, Thomas L and Suchow, Jordan W
              and Whiting, Mark E and Evans, James and Watts, Duncan J",
  journal  = "Behav. Brain Sci.",
  volume   =  47,
  pages    = "e33",
  month    =  "21~" # dec,
  year     =  2022,
  language = "en"
}

@INCOLLECTION{Hart1988-rg,
  title     = "Development of {NASA}-{TLX} (task load index): Results of
               empirical and theoretical research",
  author    = "Hart, Sandra G and Staveland, Lowell E",
  booktitle = "Advances in Psychology",
  publisher = "Elsevier",
  volume    =  382,
  pages     = "139--183",
  series    = "Advances in psychology",
  year      =  1988
}

@ARTICLE{McKee2023-hn,
  title    = "Scaffolding cooperation in human groups with deep reinforcement
              learning",
  author   = "McKee, Kevin R and Tacchetti, Andrea and Bakker, Michiel A and
              Balaguer, Jan and Campbell-Gillingham, Lucy and Everett, Richard
              and Botvinick, Matthew",
  journal  = "Nat Hum Behav",
  volume   =  7,
  number   =  10,
  pages    = "1787--1796",
  month    =  oct,
  year     =  2023,
  language = "en"
}

@ARTICLE{Mesmer-Magnus2009-fa,
  title    = "Information sharing and team performance: a meta-analysis",
  author   = "Mesmer-Magnus, Jessica R and Dechurch, Leslie A",
  journal  = "J. Appl. Psychol.",
  volume   =  94,
  number   =  2,
  pages    = "535--546",
  month    =  mar,
  year     =  2009,
  language = "en"
}

@ARTICLE{Toma2009-qm,
  title    = "Hidden profiles and concealed information: strategic information
              sharing and use in group decision making",
  author   = "Toma, Claudia and Butera, Fabrizio",
  journal  = "Pers. Soc. Psychol. Bull.",
  volume   =  35,
  number   =  6,
  pages    = "793--806",
  month    =  jun,
  year     =  2009,
  language = "en"
}

@INPROCEEDINGS{Shamekhi2019-ht,
  title     = "A Multimodal Robot-Driven Meeting Facilitation System for Group
               Decision-Making Sessions",
  author    = "Shamekhi, Ameneh and Bickmore, Timothy",
  booktitle = "2019 International Conference on Multimodal Interaction",
  publisher = "Association for Computing Machinery",
  address   = "New York, NY, USA",
  pages     = "279--290",
  series    = "ICMI '19",
  month     =  "14~" # oct,
  year      =  2019
}

@ARTICLE{Mao2024-tb,
  title         = "Multi-User Chat Assistant ({MUCA}): a Framework Using {LLMs}
                   to Facilitate Group Conversations",
  author        = "Mao, Manqing and Ting, Paishun and Xiang, Yijian and Xu,
                   Mingyang and Chen, Julia and Lin, Jianzhe",
  journal       = "arXiv [cs.CL]",
  month         =  "10~" # jan,
  year          =  2024,
  archivePrefix = "arXiv",
  primaryClass  = "cs.CL"
}

@ARTICLE{Schulz-Hardt2006-os,
  title     = "Group decision making in hidden profile situations: dissent as a
               facilitator for decision quality",
  author    = "Schulz-Hardt, Stefan and Brodbeck, Felix C and Mojzisch, Andreas
               and Kerschreiter, Rudolf and Frey, Dieter",
  journal   = "J. Pers. Soc. Psychol.",
  publisher = "American Psychological Association (APA)",
  volume    =  91,
  number    =  6,
  pages     = "1080--1093",
  month     =  dec,
  year      =  2006,
  language  = "en"
}

@ARTICLE{Noy2023-pq,
  title     = "Experimental evidence on the productivity effects of generative
               artificial intelligence",
  author    = "Noy, Shakked and Zhang, Whitney",
  journal   = "Science",
  publisher = "American Association for the Advancement of Science",
  volume    =  381,
  number    =  6654,
  pages     = "187--192",
  month     =  "14~" # jul,
  year      =  2023,
  language  = "en"
}

@INCOLLECTION{Wu2021-eb,
  title     = "{AI} creativity and the human-{AI} co-creation model",
  author    = "Wu, Zhuohao and Ji, Danwen and Yu, Kaiwen and Zeng, Xianxu and
               Wu, Dingming and Shidujaman, Mohammad",
  booktitle = "Human-Computer Interaction. Theory, Methods and Tools",
  publisher = "Springer International Publishing",
  address   = "Cham",
  pages     = "171--190",
  series    = "Lecture notes in computer science",
  year      =  2021
}

@ARTICLE{Stasser2003-kw,
  title     = "Hidden profiles: A brief history",
  author    = "Stasser, Garold and Titus, William",
  journal   = "Psychol. Inq.",
  publisher = "Informa UK Limited",
  volume    =  14,
  number    = "3-4",
  pages     = "304--313",
  month     =  oct,
  year      =  2003
}

@INPROCEEDINGS{Chiang2024-cz,
  title     = "Enhancing {AI}-assisted group decision making through
               {LLM}-powered devil's advocate",
  author    = "Chiang, Chun-Wei and Lu, Zhuoran and Li, Zhuoyan and Yin, Ming",
  booktitle = "Proceedings of the 29th International Conference on Intelligent
               User Interfaces",
  publisher = "ACM",
  address   = "New York, NY, USA",
  volume    =  1,
  month     =  "18~" # mar,
  year      =  2024
}

@ARTICLE{Gigerenzer1996-tu,
  title     = "Reasoning the fast and frugal way: Models of bounded rationality",
  author    = "Gigerenzer, Gerd and Goldstein, Daniel G",
  journal   = "Psychol. Rev.",
  publisher = "American Psychological Association (APA)",
  volume    =  103,
  number    =  4,
  pages     = "650--669",
  year      =  1996
}

@ARTICLE{Xiao2016-dp,
  title     = "Does information sharing always improve team decision making? An
               examination of the hidden profile condition in new product
               development",
  author    = "Xiao, Yazhen and Zhang, Haisu and Basadur, Timothy M",
  journal   = "J. Bus. Res.",
  publisher = "Elsevier BV",
  volume    =  69,
  number    =  2,
  pages     = "587--595",
  month     =  "9~" # feb,
  year      =  2016
}

@ARTICLE{Stasser1985-em,
  title     = "Pooling of unshared information in group decision making: Biased
               information sampling during discussion",
  author    = "Stasser, Garold and Titus, William",
  journal   = "J. Pers. Soc. Psychol.",
  publisher = "American Psychological Association (APA)",
  volume    =  48,
  number    =  6,
  pages     = "1467--1478",
  month     =  jun,
  year      =  1985,
  language  = "en"
}

@ARTICLE{Larson1996-zt,
  title     = "Diagnosing groups: Charting the flow of information in medical
               decision-making teams",
  author    = "Larson, James R and Christensen, Caryn and Abbott, Ann S and
               Franz, Timothy M",
  journal   = "J. Pers. Soc. Psychol.",
  publisher = "American Psychological Association (APA)",
  volume    =  71,
  number    =  2,
  pages     = "315--330",
  year      =  1996,
  language  = "en"
}

@ARTICLE{Latane1979-kb,
  title     = "Many hands make light the work: The causes and consequences of
               social loafing",
  author    = "Latané, Bibb and Williams, Kipling and Harkins, Stephen",
  journal   = "J. Pers. Soc. Psychol.",
  publisher = "American Psychological Association (APA)",
  volume    =  37,
  number    =  6,
  pages     = "822--832",
  month     =  jun,
  year      =  1979,
  language  = "en"
}

@ARTICLE{Schulz-Hardt2016-og,
  title     = "Preference-consistent information repetitions during discussion:
               Do they affect subsequent judgments and decisions?",
  author    = "Schulz-Hardt, Stefan and Giersiepen, Annika and Mojzisch, Andreas",
  journal   = "J. Exp. Soc. Psychol.",
  publisher = "Elsevier BV",
  volume    =  64,
  pages     = "41--49",
  month     =  may,
  year      =  2016,
  language  = "en"
}

@ARTICLE{Brodbeck2007-ae,
  title     = "Group decision making under conditions of distributed knowledge:
               The information asymmetries model",
  author    = "Brodbeck, Felix C and Kerschreiter, Rudolf and Mojzisch, Andreas
               and Schulz-Hardt, Stefan",
  journal   = "Acad. Manage. Rev.",
  publisher = "Academy of Management",
  volume    =  32,
  number    =  2,
  pages     = "459--479",
  month     =  apr,
  year      =  2007,
  language  = "en"
}

@ARTICLE{Kozlowski2006-pb,
  title     = "Enhancing the effectiveness of work groups and teams",
  author    = "Kozlowski, Steve W J and Ilgen, Daniel R",
  journal   = "Psychol. Sci. Public Interest",
  publisher = "SAGE Publications",
  volume    =  7,
  number    =  3,
  pages     = "77--124",
  month     =  dec,
  year      =  2006,
  language  = "en"
}

@ARTICLE{Wittenbaum2004-oe,
  title     = "From cooperative to motivated information sharing in groups:
               moving beyond the hidden profile paradigm",
  author    = "Wittenbaum, Gwen M and Hollingshead, Andrea B and Botero, Isabel
               C",
  journal   = "Commun. Monogr.",
  publisher = "Informa UK Limited",
  volume    =  71,
  number    =  3,
  pages     = "286--310",
  month     =  "1~" # sep,
  year      =  2004,
  language  = "en"
}

@ARTICLE{Reimer2010-no,
  title     = "Decision-making groups attenuate the discussion bias in favor of
               shared information: A meta-analysis",
  author    = "Reimer, Torsten and Reimer, Andrea and Czienskowski, Uwe",
  journal   = "Commun. Monogr.",
  publisher = "Informa UK Limited",
  volume    =  77,
  number    =  1,
  pages     = "121--142",
  month     =  "1~" # mar,
  year      =  2010,
  language  = "en"
}

@ARTICLE{Isenberg1986-tk,
  title     = "Group polarization: A critical review and meta-analysis",
  author    = "Isenberg, Daniel J",
  journal   = "J. Pers. Soc. Psychol.",
  publisher = "American Psychological Association (APA)",
  volume    =  50,
  number    =  6,
  pages     = "1141--1151",
  month     =  jun,
  year      =  1986,
  language  = "en"
}

@ARTICLE{Lee2024-jy,
  title         = "Conversational agents as catalysts for critical thinking:
                   Challenging design fixation in group design",
  author        = "Lee, Soohwan and Hwang, Seoyeong and Lee, Kyungho",
  journal       = "arXiv [cs.HC]",
  month         =  "16~" # jun,
  year          =  2024,
  archivePrefix = "arXiv",
  primaryClass  = "cs.HC"
}

@BOOK{Surowiecki2005-qv,
  title     = "The wisdom of crowds",
  author    = "Surowiecki, James",
  publisher = "Anchor Books",
  address   = "New York, NY",
  month     =  "16~" # aug,
  year      =  2005
}

@ARTICLE{Faulmuller2012-ls,
  title     = "Do you want to convince me or to be understood?
               Preference-consistent information sharing and its motivational
               determinants",
  author    = "Faulmüller, Nadira and Mojzisch, Andreas and Kerschreiter, Rudolf
               and Schulz-Hardt, Stefan",
  journal   = "Pers. Soc. Psychol. Bull.",
  publisher = "SAGE Publications",
  volume    =  38,
  number    =  12,
  pages     = "1684--1696",
  month     =  "28~" # dec,
  year      =  2012,
  language  = "en"
}

@ARTICLE{Mojzisch2010-ol,
  title     = "Biased evaluation of information during discussion: Disentangling
               the effects of preference consistency, social validation, and
               ownership of information",
  author    = "Mojzisch, Andreas and Grouneva, Lilia and Schulz-Hardt, Stefan",
  journal   = "Eur. J. Soc. Psychol.",
  publisher = "Wiley",
  volume    =  40,
  number    =  6,
  pages     = "946--956",
  month     =  "1~" # oct,
  year      =  2010,
  language  = "en"
}

@ARTICLE{Dennis1996-ge,
  title     = "Information exchange and use in group decision making: You can
               lead a group to information, but you can't make it think",
  author    = "Dennis, Alan R",
  journal   = "MIS Q",
  publisher = "JSTOR",
  volume    =  20,
  number    =  4,
  pages     =  433,
  month     =  dec,
  year      =  1996,
  language  = "en"
}

@ARTICLE{Stasser2000-yo,
  title     = "Pooling unshared information: The benefits of knowing how access
               to information is distributed among group members",
  author    = "Stasser, Garold and Vaughan, Sandra I and Stewart, Dennis D",
  journal   = "Organ. Behav. Hum. Decis. Process.",
  publisher = "Elsevier BV",
  volume    =  82,
  number    =  1,
  pages     = "102--116",
  month     =  "1~" # may,
  year      =  2000,
  language  = "en"
}

@INCOLLECTION{Baron2005-ff,
  title     = "So right it's wrong: Groupthink and the ubiquitous nature of
               polarized group decision making",
  author    = "Baron, Robert S",
  booktitle = "Advances in Experimental Social Psychology",
  publisher = "Elsevier",
  volume    =  37,
  pages     = "219--253",
  series    = "Advances in experimental social psychology",
  month     =  "1~" # jan,
  year      =  2005
}

@MISC{AmesUnknown-vy,
  title        = "Grogan Air: Team Decision Exercise",
  author       = "Ames, Daniel",
  booktitle    = "CaseWorks",
  howpublished = "\url{https://caseworks.business.columbia.edu/caseworks/grogan-air-team-decision-exercise}",
  note         = "Accessed: 2025-3-3",
  language     = "en"
}

@ARTICLE{Tessler2024-hq,
  title     = "{AI} can help humans find common ground in democratic
               deliberation",
  author    = "Tessler, Michael Henry and Bakker, Michiel A and Jarrett, Daniel
               and Sheahan, Hannah and Chadwick, Martin J and Koster, Raphael
               and Evans, Georgina and Campbell-Gillingham, Lucy and Collins,
               Tantum and Parkes, David C and Botvinick, Matthew and
               Summerfield, Christopher",
  journal   = "Science",
  publisher = "American Association for the Advancement of Science",
  volume    =  386,
  number    =  6719,
  pages     = "eadq2852",
  month     =  "18~" # oct,
  year      =  2024,
  language  = "en"
}

@ARTICLE{Kelly2004-dj,
  title     = "Time pressure and group performance: Exploring underlying
               processes in the Attentional Focus Model",
  author    = "Kelly, Janice R and Loving, Timothy J",
  journal   = "J. Exp. Soc. Psychol.",
  publisher = "Elsevier BV",
  volume    =  40,
  number    =  2,
  pages     = "185--198",
  month     =  mar,
  year      =  2004,
  language  = "en"
}

@ARTICLE{Kelly1999-wk,
  title     = "Group decision making: The effects of initial preferences and
               time pressure",
  author    = "Kelly, Janice R and Karau, Steven J",
  journal   = "Pers. Soc. Psychol. Bull.",
  publisher = "SAGE Publications",
  volume    =  25,
  number    =  11,
  pages     = "1342--1354",
  month     =  nov,
  year      =  1999,
  language  = "en"
}

@ARTICLE{Cruz1997-dy,
  title     = "The impact of group size and proportion of shared information on
               the exchange and integration of information in groups",
  author    = "Cruz, Michael G and Boster, Franklin J and Rodríguez, Jóse I",
  journal   = "Communic. Res.",
  publisher = "SAGE Publications",
  volume    =  24,
  number    =  3,
  pages     = "291--313",
  month     =  jun,
  year      =  1997,
  language  = "en"
}

@ARTICLE{Stasser1987-cy,
  title     = "Effects of information load and percentage of shared information
               on the dissemination of unshared information during group
               discussion",
  author    = "Stasser, Garold and Titus, William",
  journal   = "J. Pers. Soc. Psychol.",
  publisher = "American Psychological Association (APA)",
  volume    =  53,
  number    =  1,
  pages     = "81--93",
  month     =  jul,
  year      =  1987,
  language  = "en"
}

@ARTICLE{Shirani2006-rk,
  title     = "Sampling and pooling of decision-relevant information: Comparing
               the efficiency of face-to-face and {GSS} supported groups",
  author    = "Shirani, Ashraf I",
  journal   = "Inf. Manag.",
  publisher = "Elsevier BV",
  volume    =  43,
  number    =  4,
  pages     = "521--529",
  month     =  jun,
  year      =  2006,
  language  = "en"
}

@ARTICLE{Lam2000-pr,
  title     = "Improving group decisions by better pooling information: A
               comparative advantage of group decision support systems",
  author    = "Lam, Simon S K and Schaubroeck, John",
  journal   = "J. Appl. Psychol.",
  publisher = "American Psychological Association (APA)",
  volume    =  85,
  number    =  4,
  pages     = "565--573",
  year      =  2000,
  language  = "en"
}

@ARTICLE{Kerr2009-ws,
  title     = "Beyond brainstorming: The effectiveness of computer-mediated
               communication for convergence and negotiation tasks",
  author    = "Kerr, David S and Murthy, Uday S",
  journal   = "Int. J. Acc. Inf. Syst.",
  publisher = "Elsevier BV",
  volume    =  10,
  number    =  4,
  pages     = "245--262",
  month     =  "1~" # dec,
  year      =  2009,
  language  = "en"
}

@ARTICLE{Brodbeck2002-in,
  title     = "The dissemination of critical, unshared information in
               decision‐making groups: the effects of pre‐discussion dissent",
  author    = "Brodbeck, Felix C and Kerschreiter, Rudolf and Mojzisch, Andreas
               and Frey, Dieter and Schulz-Hardt, Stefan",
  journal   = "Eur. J. Soc. Psychol.",
  publisher = "Wiley",
  volume    =  32,
  number    =  1,
  pages     = "35--56",
  month     =  jan,
  year      =  2002,
  language  = "en"
}

@ARTICLE{Stasser1995-gs,
  title     = "Expert roles and information exchange during discussion: The
               importance of knowing who knows what",
  author    = "Stasser, Garold and Stewart, Dennis D and Wittenbaum, Gwen M",
  journal   = "J. Exp. Soc. Psychol.",
  publisher = "Elsevier BV",
  volume    =  31,
  number    =  3,
  pages     = "244--265",
  month     =  may,
  year      =  1995,
  language  = "en"
}

@ARTICLE{Small2023-qg,
  title         = "Opportunities and risks of {LLMs} for scalable deliberation
                   with Polis",
  author        = "Small, Christopher T and Vendrov, Ivan and Durmus, Esin and
                   Homaei, Hadjar and Barry, Elizabeth and Cornebise, Julien and
                   Suzman, Ted and Ganguli, Deep and Megill, Colin",
  journal       = "arXiv [cs.SI]",
  month         =  "20~" # jun,
  year          =  2023,
  archivePrefix = "arXiv",
  primaryClass  = "cs.SI"
}

@ARTICLE{Phillips1993-fb,
  title     = "Faciliated work groups: Theory and practice",
  author    = "Phillips, Lawrence D and Phillips, Maryann C",
  journal   = "J. Oper. Res. Soc.",
  publisher = "JSTOR",
  volume    =  44,
  number    =  6,
  pages     =  533,
  month     =  jun,
  year      =  1993,
  language  = "en"
}

@ARTICLE{Li2022-gb,
  title     = "Heterogeneous large-scale group decision making using fuzzy
               cluster analysis and its application to emergency response plan
               selection",
  author    = "Li, Guangxu and Kou, Gang and Peng, Yi",
  journal   = "IEEE Trans. Syst. Man Cybern. Syst.",
  publisher = "Institute of Electrical and Electronics Engineers (IEEE)",
  volume    =  52,
  number    =  6,
  pages     = "3391--3403",
  month     =  jun,
  year      =  2022
}

@ARTICLE{DiPierro2022-lw,
  title     = "Groupthink among health professional teams in patient care: A
               scoping review",
  author    = "DiPierro, Karissa and Lee, Hannah and Pain, Kevin J and Durning,
               Steven J and Choi, Justin J",
  journal   = "Med. Teach.",
  publisher = "Informa UK Limited",
  volume    =  44,
  number    =  3,
  pages     = "309--318",
  month     =  mar,
  year      =  2022,
  language  = "en"
}

@BOOK{Tavana1993-pq,
  title     = "An {AHP}-Delphi group decision support system applied to conflict
               resolution in hiring decisions",
  author    = "Tavana, Madjid and Kennedy, Dennis T and Rappaport, Jack and
               Ugras, Yusuf Joseph",
  publisher = "Journal of Personality and Social Psychology",
  year      =  1993
}

@INCOLLECTION{Hackman1975-yi,
  title     = "Group tasks, group interaction process, and group performance
               effectiveness: A review and proposed integration",
  author    = "Hackman, J Richard and Morris, Charles G",
  booktitle = "Advances in Experimental Social Psychology",
  publisher = "Elsevier",
  volume    =  8,
  pages     = "45--99",
  series    = "Advances in experimental social psychology",
  month     =  "1~" # jan,
  year      =  1975
}

@BOOK{McGrath1984-gp,
  title     = "Groups: Interaction and Performance",
  author    = "McGrath, Joseph E",
  publisher = "Prentice Hall",
  address   = "Old Tappan, NJ",
  month     =  "1~" # jan,
  year      =  1984,
  language  = "en"
}

@ARTICLE{McCROSKEY1977-sm,
  title     = "Oral communication apprehension: A summary of recent theory and
               research",
  author    = "McCROSKEY, James C",
  journal   = "Hum. Commun. Res.",
  publisher = "Oxford University Press (OUP)",
  volume    =  4,
  number    =  1,
  pages     = "78--96",
  month     =  "1~" # sep,
  year      =  1977,
  language  = "en"
}

@ARTICLE{Kim2021-qw,
  title     = "Moderator chatbot for deliberative discussion: Effects of
               discussion structure and discussant facilitation",
  author    = "Kim, Soomin and Eun, Jinsu and Seering, Joseph and Lee, Joonhwan",
  journal   = "Proc. ACM Hum. Comput. Interact.",
  publisher = "Association for Computing Machinery (ACM)",
  volume    =  5,
  number    = "CSCW1",
  pages     = "1--26",
  month     =  "13~" # apr,
  year      =  2021,
  language  = "en"
}

@ARTICLE{Fink1984-bx,
  title     = "Consensus methods: characteristics and guidelines for use",
  author    = "Fink, A and Kosecoff, J and Chassin, M and Brook, R H",
  journal   = "Am. J. Public Health",
  publisher = "American Public Health Association",
  volume    =  74,
  number    =  9,
  pages     = "979--983",
  month     =  sep,
  year      =  1984,
  language  = "en"
}

@INPROCEEDINGS{Claggett2025-ov,
  title     = "Relational {AI}: Facilitating intergroup cooperation with
               socially aware conversational support",
  author    = "Claggett, Elijah L and Kraut, Robert E and Shirado, Hirokazu",
  booktitle = "Proceedings of the 2025 CHI Conference on Human Factors in
               Computing Systems",
  publisher = "ACM",
  address   = "New York, NY, USA",
  pages     = "1--22",
  month     =  "26~" # apr,
  year      =  2025,
  language  = "en"
}

@ARTICLE{Benjamini1995-tq,
  title     = "Controlling the false discovery rate: a practical and powerful
               approach to multiple testing",
  author    = "Benjamini, Y and Hochberg, Y",
  journal   = "Journal of the royal statistical society series b-methodological",
  publisher = "[Royal Statistical Society, Oxford University Press]",
  volume    =  57,
  number    =  1,
  pages     = "289--300",
  year      =  1995,
  language  = "en"
}

@INPROCEEDINGS{Chen2025-je,
  title     = "Are we on track? {AI}-assisted active and passive goal reflection
               during meetings",
  author    = "Chen, Xinyue and Tankelevitch, Lev and Vanukuru, Rishi and Scott,
               Ava Elizabeth and Panda, Payod and Rintel, Sean",
  booktitle = "Proceedings of the 2025 CHI Conference on Human Factors in
               Computing Systems",
  publisher = "ACM",
  address   = "New York, NY, USA",
  pages     = "1--22",
  month     =  "26~" # apr,
  year      =  2025,
  language  = "en"
}

@ARTICLE{Qian2025-eu,
  title         = "Deliberate Lab: A platform for real-time human-{AI} social
                   experiments",
  author        = "Qian, Crystal and Tsai, Vivian and Behr, Michael and Hussein,
                   Nada and Laugier, Léo and Thain, Nithum and Dixon, Lucas",
  journal       = "arXiv [cs.HC]",
  month         =  "14~" # oct,
  year          =  2025,
  archivePrefix = "arXiv",
  primaryClass  = "cs.HC"
}

@INPROCEEDINGS{Vanukuru2025-od,
  title     = "Designing interfaces that support temporal work across meetings
               with generative {AI}",
  author    = "Vanukuru, Rishi and Panda, Payod and Chen, Xinyue and Scott, Ava
               Elizabeth and Tankelevitch, Lev and Rintel, Sean",
  booktitle = "Proceedings of the 2025 ACM Designing Interactive Systems
               Conference",
  publisher = "ACM",
  address   = "New York, NY, USA",
  pages     = "3600--3620",
  month     =  "5~" # jul,
  year      =  2025
}

@INPROCEEDINGS{Houtti2025-eh,
  title     = "Observe, ask, intervene: Designing {AI} agents for more inclusive
               meetings",
  author    = "Houtti, Mo and Zhou, Moyan and Terveen, Loren and Chancellor,
               Stevie",
  booktitle = "Proceedings of the 2025 CHI Conference on Human Factors in
               Computing Systems",
  publisher = "ACM",
  address   = "New York, NY, USA",
  pages     = "1--18",
  month     =  "26~" # apr,
  year      =  2025
}

@INPROCEEDINGS{Ma2025-aj,
  title     = "Towards human-{AI} deliberation: Design and evaluation of
               {LLM}-empowered deliberative {AI} for {AI}-assisted
               decision-making",
  author    = "Ma, Shuai and Chen, Qiaoyi and Wang, Xinru and Zheng, Chengbo and
               Peng, Zhenhui and Yin, Ming and Ma, Xiaojuan",
  booktitle = "Proceedings of the 2025 CHI Conference on Human Factors in
               Computing Systems",
  publisher = "ACM",
  address   = "New York, NY, USA",
  pages     = "1--23",
  month     =  "26~" # apr,
  year      =  2025,
  language  = "en"
}

@ARTICLE{Miranda1999-bd,
  title     = "Meeting Facilitation: Process Versus Content Interventions",
  author    = "Miranda, Shalla M and Bostrom, Robert P",
  journal   = "J. Manag. Inf. Syst.",
  publisher = "Informa UK Limited",
  volume    =  15,
  number    =  4,
  pages     = "89--114",
  month     =  mar,
  year      =  1999
}

@ARTICLE{DeSanctis1987-oh,
  title     = "A foundation for the study of group decision support systems",
  author    = "DeSanctis, Gerardine and Gallupe, R Brent",
  journal   = "Manage. Sci.",
  publisher = "Institute for Operations Research and the Management Sciences
               (INFORMS)",
  volume    =  33,
  number    =  5,
  pages     = "589--609",
  month     =  may,
  year      =  1987,
  language  = "en"
}

@ARTICLE{Diederich2022-yj,
  title     = "On the design of and interaction with conversational agents: An
               organizing and assessing review of human-computer interaction
               research",
  author    = "Diederich, Stephan and {University of Göttingen, Germany} and
               Brendel, Alfred Benedikt and Morana, Stefan and Kolbe, Lutz and
               {TU Dresden, Germany} and {Saarland University, Germany} and
               {University of Göttingen, Germany}",
  journal   = "J. Assoc. Inf. Syst.",
  publisher = "Association for Information Systems",
  volume    =  23,
  number    =  1,
  pages     = "96--138",
  year      =  2022,
  language  = "en"
}

@ARTICLE{Papachristou2025-ex,
  title     = "Leveraging Large Language Models for collective decision-making",
  author    = "Papachristou, Marios and Yang, Longqi and Hsu, Chin-Chia",
  journal   = "Proc. ACM Hum. Comput. Interact.",
  publisher = "Association for Computing Machinery (ACM)",
  volume    =  9,
  number    =  7,
  pages     = "1--44",
  month     =  "18~" # oct,
  year      =  2025,
  language  = "en"
}

@ARTICLE{Mozannar2024-qh,
  title         = "The {RealHumanEval}: Evaluating large language models'
                   abilities to support programmers",
  author        = "Mozannar, Hussein and Chen, Valerie and Alsobay, Mohammed and
                   Das, Subhro and Zhao, Sebastian and Wei, Dennis and
                   Nagireddy, Manish and Sattigeri, Prasanna and Talwalkar,
                   Ameet and Sontag, David",
  journal       = "Transact. Mach. Learn. Res.",
  month         =  jan,
  year          =  2025,
  archivePrefix = "arXiv",
  primaryClass  = "cs.SE",
  language      = "en"
}

@INPROCEEDINGS{Kumar2023-uc,
  title     = "Math education with large language models: Peril or promise?",
  author    = "Kumar, Harsh and Rothschild, David M and Goldstein, Daniel G and
               Hofman, Jake M",
  booktitle = "Artificial intelligence in education",
  publisher = "Springer Nature Switzerland",
  pages     = "60--75",
  series    = "Lecture Notes in Computer Science",
  year      =  2025,
  language  = "en"
}

@ARTICLE{Argyle2023-st,
  title     = "Leveraging {AI} for democratic discourse: Chat interventions can
               improve online political conversations at scale",
  author    = "Argyle, Lisa P and Bail, Christopher A and Busby, Ethan C and
               Gubler, Joshua R and Howe, Thomas and Rytting, Christopher and
               Sorensen, Taylor and Wingate, David",
  journal   = "Proc. Natl. Acad. Sci. U. S. A.",
  publisher = "National Academy of Sciences",
  volume    =  120,
  number    =  41,
  pages     = "e2311627120",
  month     =  "10~" # oct,
  year      =  2023,
  language  = "en"
}

@ARTICLE{Woolley_undated-rg,
  title     = "Evidence for a collective intelligence factor in the performance
               of human groups",
  author    = "Woolley, Anita Williams and Chabris, Christopher F and Pentland,
               Alex and Hashmi, Nada and Malone, Thomas W",
  journal   = "Science",
  publisher = "American Association for the Advancement of Science (AAAS)",
  volume    =  330,
  number    =  6004,
  pages     = "686--688",
  month     =  "29~" # oct,
  year      =  2010,
  language  = "en"
}

@ARTICLE{Pescetelli2021-nk,
  title     = "A variational-autoencoder approach to solve the hidden profile
               task in hybrid human-machine teams",
  author    = "Pescetelli, Niccolo and Reichert, Patrik and Rutherford, Alex",
  journal   = "PLoS One",
  publisher = "Public Library of Science (PLoS)",
  volume    =  17,
  number    =  8,
  pages     = "e0272168",
  month     =  "2~" # aug,
  year      =  2022,
  language  = "en"
}
